\documentclass[journal, draftcls, 11pt, onecolumn]{IEEEtran}

\usepackage{epsfig}
\usepackage{graphicx}
\usepackage{amsmath, amssymb}
\usepackage[ruled, vlined]{algorithm2e}
\usepackage{fancybox}
\usepackage{cite}
\usepackage{subfigure}

\newcommand{\SNR}{\mbox{\sffamily\upshape{\scriptsize{SNR}}}}

\newtheorem{theorem}{Theorem}
\newtheorem{lemma}{Lemma}

\begin{document}

\title{Information-theoretically Secret Key Generation for Fading Wireless Channels}
\author{
Chunxuan Ye$^\dag$,\ %
Suhas Mathur$^\ddag$,\ %
Alex Reznik$^\dag$,\ %
Yogendra Shah$^\dag$,\ %
Wade Trappe$^\ddag$%
\,and %
Narayan Mandayam$^\ddag$ %
\thanks{Manuscript first submitted to the IEEE Transactions on Information Forensics and Security on 23 February, 2009.}
\thanks{$^\dag$ InterDigital Communications, LLC, King of Prussia, PA 19406, USA.}
\thanks{E-mail:\texttt{\{chunxuan.ye, alex.reznik,
yogendra.shah\}@interdigital.com} }%
\thanks{$^\ddag$ WINLAB, Rutgers University, 671 Route 1 South, North Brunswick, NJ 08902,
USA.}
\thanks{E-mail: \texttt{\{suhas, trappe, narayan\}@winlab.rutgers.edu} }%
\thanks{S. Mathur, W. Trappe and N. Mandayam are supported in part by by NSF grant CNS-0626439 and DARPA grant W31P4Q-07-1-002}
\thanks{Portions of this work have been previous presented at the \emph{IEEE International Symposium on Information Theory}, Seattle, WA, July 2006 and \emph{ACM Conference on Mobile Computing and Networking}, San Francisco, CA, Sept. 2008.}
}

\maketitle

 \begin{abstract}
The multipath-rich wireless environment associated with typical wireless usage scenarios is characterized by a fading channel response that is time-varying, location-sensitive, and uniquely shared by a given transmitter-receiver pair. The complexity associated with a richly scattering environment implies that the short-term fading process is inherently hard to predict and best modeled stochastically, with rapid decorrelation properties in space, time and frequency. In this paper, we demonstrate how the channel state between a wireless transmitter and receiver can be used as the basis for building practical secret key generation protocols between two entities. We begin by presenting a scheme based on level crossings of the fading process, which is well-suited for the Rayleigh and Rician fading models associated with a richly scattering environment. Our level crossing algorithm is simple, and incorporates a self-authenticating mechanism to prevent adversarial manipulation of message exchanges during the protocol. Since the level crossing algorithm is best suited for fading processes that exhibit symmetry in their underlying distribution, we present a second and more powerful approach that is suited for more general channel state distributions. This second approach is motivated by observations from quantizing jointly Gaussian processes, but exploits empirical measurements to set quantization boundaries and a heuristic log likelihood ratio estimate to achieve an improved secret key generation rate.  We validate both proposed protocols through experimentations using a customized 802.11a platform, and show for the typical WiFi channel that reliable secret key establishment can be accomplished at rates on the order of 10 bits/second.
\end{abstract}

\section{Introduction}

The problem of secret key generation from correlated information was
first studied by Maurer \cite{Mau93}, and Ahlswede and Csisz\'ar
\cite{AhlCsi93}. In a basic secret key generation problem, called
the {\it basic source model}, two legitimate terminals (Alice and
Bob)\footnote{Unless otherwise specified, all the terminals in this
paper refer to legitimate terminals, and hence the term
``legitimate'' will be omitted henceforth.} observe a common random
source that is inaccessible to an eavesdropper. Modeling the
observations as memoryless, we can define the model as follows:
Alice and Bob respectively observe $n$ independent and identically
distributed (i.i.d.) repetitions of the dependent random variables
$X$ and $Y$, denoted by $X^n=(X_1,\cdots, X_n)$ and $Y^n=(Y_1,
\cdots , Y_n)$. In any given time instance, the observation pair
$(X_i, Y_i)$ is highly statistically dependent. Based on their
dependent observations, Alice and Bob generate a common secret key
by communicating over a public error-free channel, with the
communication denoted collectively by ${\bf V}$.

A random variable $K$ with finite range ${\cal K}$ represents an
{\it $\varepsilon$-secret key} for Alice and Bob, achievable with
communication ${\bf V}$, if there exist two functions $f_A$, $f_B$
such that $K_A=f_A(X^n, {\bf V})$, $K_B=f_B(Y^n, {\bf V})$, and for
any $\varepsilon>0$,
\begin{equation}
\Pr(K=K_A=K_B)\geq 1-\varepsilon,\label{e2.1}
\end{equation}
\begin{equation}
I(K; {\bf V})\leq \varepsilon, \label{e2.2}
\end{equation}
\begin{equation}
H(K) \geq \log |{\cal K}|-\varepsilon. \label{e2.3}
\end{equation}
Here, condition (\ref{e2.1}) ensures that Alice and Bob generate the
same secret key with high probability; condition (\ref{e2.2})
ensures such secret key is effectively concealed from the
eavesdropper observing the public communication ${\bf V}$; and
condition (\ref{e2.3}) ensures such a secret key is nearly uniformly
distributed.

An achievable secret key rate $R$ is defined \cite{Mau93},
\cite{AhlCsi93} to be a value such that for every $\varepsilon>0$
and sufficiently large $n$, an $\varepsilon$-secret key $K$ is
achievable with suitable communication such that
$\frac{1}{n}H(K)\geq R-\varepsilon$.  The supremum of all achievable
secret key rates is the {\it secret key capacity} denoted by
$C_{SK}$. For the model presented above, this is given by
\cite{Mau93}, \cite{AhlCsi93}, \cite{Mau94}, \cite{MauWol00}
\begin{equation}
 C_{SK} = I(X;Y). \label{e2.3a}
\end{equation}
This result holds for both discrete and continuous random variables
$X$ and $Y$, as long as $I(X; Y)$ is finite (cf. \cite{YeRez06},
\cite{ Nit08}).

The model defined above assumes the eavesdropper (i.e. Eve) may
observe the transmissions on the public channel, but is unable to
tamper with them and has no access to any other useful side
information.  The case of an eavesdropper with access to side
information has received significant attention (see, e.g.,
\cite{Mau93}, \cite{AhlCsi93}, \cite{RSW_IT03}, \cite{MRW_IT03});
unfortunately the capacity problem remains open in this case. The
case of an eavesdropper with the ability to tamper with the
transmissions on the public channel has been addressed in a
comprehensive analysis by Maurer and Wolf \cite{Mau97},
\cite{MauWol03a}, \cite{MauWol03b}, \cite{MauWol03c}.

A practical implementation of secret-key agreement schemes follows a
basic 3-phase protocol defined by Maurer \emph{et.al.}. The first
phase, {\it advantage distillation} \cite{Mau93}, \cite{CacMau97a},
is aimed at providing two terminals an advantage over the
eavesdropper when the eavesdropper has access to side information.
We do not consider this scenario (as we shall see shortly, it is not
necessary for secrecy generation from wireless channels) and,
therefore, do not address {\it advantage distillation}.

The second phase, {\it information reconciliation} \cite{BenBra86},
\cite{BenBes92}, \cite{BraSal94}, is aimed at generating an
identical random sequence between the two terminals by exploiting
the public channel. For a better secret key rate, the entropy of
this random sequence should be maximized, while the amount of
information transmitted on the public channel should be minimized.
This suggests an innate connection between the information
reconciliation phase of the secrecy agreement protocol and
Slepian-Wolf data compression. This connection was formalized by
\cite{CsiNar04} in the general setting of multi-terminal secrecy
generation.

The connection between secrecy generation and data compression is of
significant practical, as well as theoretical interest. Considering
the duality between Slepian-Wolf data compression and channel coding
(e.g., \cite{GarZha01}, \cite{LivXio02}, \cite{PraRam03},
\cite{ColLee06}, \cite{CheHe08}, etc), the relationship between
secrecy generation and data compression allows capacity-achieving
channel codes, like Turbo codes or LDPC codes, to be used for the
information reconciliation phase. Moreover, the capacity-achieving
capabilities of such codes in the channel coding sense carry over to
the secrecy generation problem. A comprehensive treatment of the
application and optimality of such codes to the secrecy generation
problem can be found in \cite{BloTha06}, \cite{BloBar08}.

The last phase of Maurer's protocol, {\it privacy amplification}
\cite{BenBra88}, \cite{BenBra95}, extracts a secret key from the
identical random sequence agreed to by two terminals in the
information reconciliation phase. This can be implemented by linear
mapping and universal hashing \cite{CarWeg79}, \cite{WegCar81},
\cite{BenBra95}, \cite{MauWol03c}, or by an extractor
\cite{RazRei99}, \cite{MauWol03c}, \cite{DodKat06}, \cite{DodOst08},
\cite{CraDod08}. The combination of the information reconciliation
phase and the privacy amplification phase has been considered in
\cite{CacMau97a}, \cite{YeNar05a}.

Perhaps the first practical application of the basic source model is
quantum cryptography (cf. e.g., \cite{BenBra92},\cite{NieChu00}),
where non-orthogonal states of a quantum system provide two
terminals correlated observations of randomness which are at least
partially secret from a potential eavesdropper. Quantum key
distribution schemes based on continuous random variables have been
discussed in \cite{GroVan03}, \cite{VanCar04}, \cite{BloTha06},
\cite{LodBlo07}. Less realized is the fact that wireless fading
channel provides another source \cite{HerHas95}, \cite{YeRez06},
\cite{BloBar08} of secrecy which can be used to generate
information-theoretically secure keys.  Because the source model for
secrecy establishment essentially requires \emph{a priori} existence
of a ``dirty secret'' which is then just cleaned up, such sources of
secrecy are hard to find.  To our knowledge no such sources other
than quantum entanglement and wireless channel reciprocity have been
identified to date. Further, we note that although there have been
several implementations of quantum cryptographic key establishment,
little work has been done to provide a system validation of this
process for wireless channels. This paper examines both theoretical
and practical aspects of key establishment using wireless channels
and represents one of the first validation efforts to this effect.

An alternative approach to secrecy generation from wireless channels
is based on the wiretap channel models, see e.g.\cite{BloBar08}.
However, this approach suffers from a need to make certain
assumptions as part of the security model that are hard to satisfy
in practice and has not, to date, led to a practical implementation.

A (narrowband) wireless channel is well modeled as a flat fading
channel. The fading coefficient changes in time, but the change is
rather slow (on the order of 1 msec to 1 sec, depending on terminal
velocities and other factors). For simplicity, let us consider
frequency flat fading. Roughly speaking, for a fixed time and
location, the transmitted signal $t$ and the received signal $r$ are
related via $r=Ft+Z$, where $F$ is the channel fading coefficient
and $Z$ is the additive independent noise. If the transmitted signal
$t$ is known at the receiver beforehand, (e.g., it is a training
sequence) then the receiver is able to obtain a noisy estimate of
the fading coefficient $F$. Furthermore, if both terminals send the
training sequence at approximately the same time (more precisely,
well within one channel coherence time of each other), then they can
obtain channel estimates that are highly correlated due to channel
reciprocity.  This suggests the following model: let the random
variables $X$ and $Y$ be defined by $ X=F+Z_A $, $ Y=F+Z_B $, where
$F$, $Z_A$, $Z_B$ are three independent random variables.

In data communications application, it is common to model the
channel as Rayleigh or Rician, in which case, $F$, $Z_A$ and $Z_B$
are Gaussian. Let these be distributed as ${\cal N}(0,P)$, ${\cal
N}(0,N_A)$ and ${\cal N}(0,N_B)$ respectively. A simple calculation
shows that the secret key capacity \cite{YeRez06} of this jointly
Gaussian model is
\begin{equation}
C_{SK}= \log_2\left(1+\frac{P}{N_A+N_B+\frac{N_AN_B}{P}}\right)\
bits/sample. \label{e2.6}
\end{equation}
If we let $N_A= N_B =N$ in this setting, then we get a natural
definition of SNR as $\SNR = \frac{P}{N}$, and the above secret key
capacity reduces to $\log_2\left(1+\frac{SNR}{2+1/SNR}\right)$
bits/sample.

As noted, the above calculation is relevant for the traditional Rayleigh or Rician fading model, and serves as an upper bound on the secret key establishment rate, but does not provide insight into how one can practically extract such secret bits from the underlying fading process. In this paper, we examine two different approaches for secrecy extraction from the channel state between a transmitter and receiver in a richly scattering wireless environment. Our first approach, which is based on level-crossings, is a simple algorithm that is well-suited for environments that can be characterized as Rayleigh or Rician. However, we recognize that such a method might not apply to other, general fading cases.  One way to address this problem is to consider more complex fading distribution models, such as those appropriate for ultrawideband
channels.  This has been addressed in a previous work by Wilson {\it
et. al} \cite{WilTse07} (see also \cite{AonHig05}, \cite{ImaKob06},
\cite{AziKia07}). However, we take a different approach in this paper. Inspired by our
prior work on Gaussian-based approaches, we propose a {\em universal}
reconciliation approaches for wireless channels.  This second, and more powerful method, only assumes that the channel impulse responses (CIRs)
measured at both terminals are highly correlated, and their
measurement noise is very low. Whereas the first of our two approaches
was simple,  and able to achieve a limited secret key establishment rate, our second approach is more complex, but is able to
take better advantage of the secrecy capabilities offered by CIR
measurements, which tend to have high SNR (due to a high processing
gain associated with such measurements in modern communication
systems).

In both of these cases, our goal is to come up with a practical approach to secrecy
generation from wireless channel measurements. In particular, because the statistics of the real channel
sources we utilize are not known (and that is the major challenge we believe addressed by our work),
it is impossible to make any quantitative statements about optimality of our approaches. Nevertheless,
we do want to make sure that our solution is based on solid theoretical foundation. To do so, we include
discussion of the motivating algorithms and their performance in idealized models when necessary.

Several previous attempts to use wireless channels for encrypting
communications have been proposed.  Notably, \cite{KooHas00}
exploited reciprocity of a wireless channel for secure data
transformation; \cite{HasSta96} discussed a secrecy extraction
scheme based on the phase information of received signals; the
application of the reciprocity of a wireless channel for terminal
authentication purpose was studied in \cite{PatKas07},
\cite{XiaGre07}, \cite{XiaGre08}, etc. Unlike these and other
approaches, our approach for direct secrecy generation allows the
key generation component to become a ``black box'' within a larger
communication system. Its output (a secret bit stream) can then be
used within the communication system for various purposes. This is
important, as the key generation rate is likely to be quite low, and
thus direct encryption of data will either severely limit throughput
(to less than 1 kbps in indoor channels) or result in extremely weak
secrecy.

The adversary model assumed in this paper focuses mainly on passive attacks. We do not consider authentication attacks, such as the man-in-the-middle attack, since these require an explicit
authentication mechanism between Alice and Bob and cannot be
addressed by key-extraction alone. The starting point for algorithms
presented in this paper is the successive probing of the wireless
channel by the terminals that wish to extract a secret key. Implicitly, we assume that the adversary is not engaging in an active attack against the probing process, though we note that physical layer authentication techniques, such as presented in \cite{XiaGre08} might be applicable in such an adversarial setting.  The
infeasibility of passive eavesdropping attacks on the key generation procedures is
based on the rapid spatial decorrelation of the wireless channel.
We demonstrate this using empirically computed mutual information
from the channel-probing stage, between the signals received at Bob
and Eve and comparing it with the mutual information between the
signals received at Alice and Bob. Beyond the basic eavesdropping attack, we do consider a particular type of active attack in our level-crossing algorithm in Section II, where the adversary attempts to disrupt the key extraction protocol by replacing or altering the protocol messages. In this case, we provide a method to deal with this type of active attack by cleverly using the shared fading process between Alice and Bob.

One of the goals of our work is to demonstrate
that secrecy generation can be accomplished in real-time over
real channels (and not simulation models) and in real communication systems.
To that end, results based on implementations on actual wireless platforms
(a modified commercial 802.11 a/g implementation platform) and using
over-the-air protocols are presented.  To accomplish this,
we had to work with several severe limitations
of the \emph{experimental system} at our disposal.  Consequently certain parameters
(e.g. code block length) had to be selected to be somewhat below what
they should be for a well-designed system. This, however,
does not reflect on the feasibility of proper implementation in
a system with these features designed in.  For example, nothing
would prevent a design with the code block length sufficiently long
to guarantee desired performance.  On the contrary, we believe
the demonstration of a practical implementation to be one of the major
contributions of our work.

The rest of this paper is organized as follows. Section II discusses
the simpler of our algorithms based on level crossings.  Section III
presents a more complex and more powerful approach to extracting secret bits from the channel response, as well as
some new results on secrecy generation for Gaussian sources which
motivate our solution. We conclude the paper with some final remarks
in Section IV.

\section{Level Crossing Secret Key Generation System}

In this section we describe a simple and lightweight algorithm in
\cite{MatTra08} for extracting secret bits from the wireless channel
that does not explicitly involve the use of coding techniques. While this comes at the expense of a lower secret key
rate, it reduces the complexity of the system and it still provides
a sufficiently good rate in typical indoor environments. The
algorithm uses excursions in the fading channel for generating bits
and the timing of excursions for reconciliation. Further, the system does not require i.i.d.
inputs and, therefore, does not require knowledge of the channel
coherence time a priori. We refer to this secret key generation
system as the {\it level crossing system}. We evaluate the
performance of the level crossing system and test it using
customized 802.11 hardware.

\subsection{System and Algorithm Description}

Let $F(t)$ be a stochastic process corresponding to a time-varying
parameter $F$ that describes the wireless channel shared by, and
unique to Alice and Bob. Alice and Bob transmit a known signal (a
probe) to one another in quick succession in order to derive
correlated estimates of the parameter $F$, using the received signal
by exploiting reciprocity of the wireless link.  Let $X$ and $Y$
denote the (noisy) estimates of the parameter $F$ obtained by Alice
and Bob respectively.

Alice and Bob generate a sequence of $n$ correlated estimates
$\hat{X}^n = (\hat{X}_1, \hat{X}_2, \ldots , \hat{X}_n)$ and
$\hat{Y}^n = (\hat{Y}_1, \hat{Y}_2, \ldots , \hat{Y}_n)$,
respectively, by probing the channel repeatedly in a time division
duplex (TDD) manner. Note however, that $\hat{X}_i$ (and
$\hat{Y}_i$) are no longer i.i.d. for $i=1, \ldots n$ since the
channel may be strongly correlated between successive channel
estimates.

Alice and Bob first low-pass filter their sequence of channel
estimates, $\hat{X}^n$ and $\hat{Y}^n$ respectively, by subtracting
a windowed moving average. This removes the dependence of the
channel estimates on large-scale shadow fading changes and leaves
only the small scale fading variations (see Figure \ref{IDs2}). The
resulting sequences, ${X}^n$ and ${Y}^n$ have approximately zero
mean and contain excursions in positive and negative directions with
respect to the mean. The subtraction of the windowed mean ensures
that the level-crossing algorithm below does not output long strings
of ones or zeros and that the bias towards one type of bit is
removed. The filtered sequences are then used by Alice and Bob to
build a 1-bit quantizer $\psi^u(\cdot)$ quantizer based on the
scalars $q^u_+$ and $q^u_-$ that serve as threshold levels for the
quantizer:
\begin{eqnarray} \label{q_def_u}
 q_{+}^u &=& mean(U^n) + \alpha\cdot\sigma(U^n) \\ \label{q_def_u2}
q_{-}^u &=& mean(U^n) - \alpha\cdot\sigma(U^n),
\end{eqnarray}
where the sequence $U^n = {X}^n$ for Alice and  $U^n = {Y}^n$ for
Bob. $\sigma(\cdot)$ is the standard deviation and the factor
$\alpha$ can be selected to control the quantizer thresholds. The
sequences ${X}^n$ and ${Y}^n$ are then fed into the following
locally-computed quantizer at Alice and Bob respectively:
\begin{equation}
\psi^u(x) = \left\{ \begin{array}{ll}
 1 & \textrm{if $x>q^u_{+}$}\\
 0 & \textrm{if $x<q^u_{-}$} \\
e & \textrm{Otherwise}
  \end{array}  \right.\nonumber
\end{equation}
where $e$ represents an undefined state. The superscript $u$ stands
for \emph{user} and may refer to either Alice, in which case the
quantizer function is $\psi^A(\cdot)$, or to Bob, for which the
quantizer is $\psi^B(\cdot)$. This quantizer forms the basis for
quantizing positive and negative excursions. Values between $q_-^u$
and $q_+^u$ are not assigned a bit.

It is assumed that the number $n$ of channel observations is
sufficiently large before using the level crossing system, and that
the $i^{th}$ element $X_i$ and $Y_i$ correspond to successive probes
sent by Bob and Alice respectively, for each $i = 1, \ldots, n$. The
level crossing algorithm consists of the following steps:
\begin{enumerate}
\item Alice parses the vector $X^n$ containing her
filtered channel estimates to find instances where  $m$ or more
successive estimates lie in an excursion above $q_{+}$ or below
$q_{-}$. Here, $m$ is a parameter used to denote the minimum number
of channel estimates in an excursion.
\item Alice selects a random subset of the excursions found in step 1
and, for each selected excursion, she sends Bob the index of the
channel estimate lying in the center of the excursion, as a list
$L$. Therefore, if $X_i > q_{+}$ or $< q_{-}$ for some $i=i_{start},
\ldots, i_{end}$, then she sends Bob the index
$i_{center}=\lfloor\frac{i_{start} + i_{end}}{2} \rfloor$.
\item To make sure the $L$-message received is from Alice, Bob computes
the fraction of indices in $L$ where $Y^n$ lies in an excursion
spanning $(m-1)$ or more estimates. If this fraction is less than
$\frac{1}{2} + \epsilon$, for some fixed parameter $0 < \epsilon <
\frac{1}{2}$, Bob concludes that the message was not sent by Alice,
implying an adversary has injected a fake $L$-message.
\item If the check above passes, Bob replies to Alice with a message
$\tilde{L}$ containing those indices in $L$ at which $Y^n$ lies in
an excursion. Bob computes $K_B = \psi^B(Y_i; i\in \tilde{L})$ to
obtain $N$ bits. The first $N_{au}$ bits are used as an
authentication key to compute a message authentication code (MAC) of
$\tilde{L}$. The remaining $N-N_{au}$ bits are kept as the extracted
secret key. The overall message sent by Bob is $\left\{\tilde{L},
MAC\left(K_{au},\tilde{L}\right)\right\}$. Practical
implementations, for example, one could use CBC-MAC as the implementation
for MAC, and use a key $K_{au}$ of length $N_{au} = 128$ bits.
\item Upon receiving this message from Bob, Alice uses $\tilde{L}$ to
form the sequence of bits $K_A = \psi^A(X_i; i\in \tilde{L})$. She
uses the first $N_{au}$ bits of $K_A$ as the authentication key
$K_{au} = K_A(1, \ldots, N_{au})$, and, using $K_{au}$, she verifies
the MAC to confirm that the package was indeed sent by Bob. Since
Eve does not know the bits in $K_{au}$ generated by Bob, she cannot
modify the $\tilde{L}$-message without failing the MAC verification
at Alice.
\end{enumerate}

\begin{figure}
 \begin{center}
  \includegraphics[scale= 0.5]{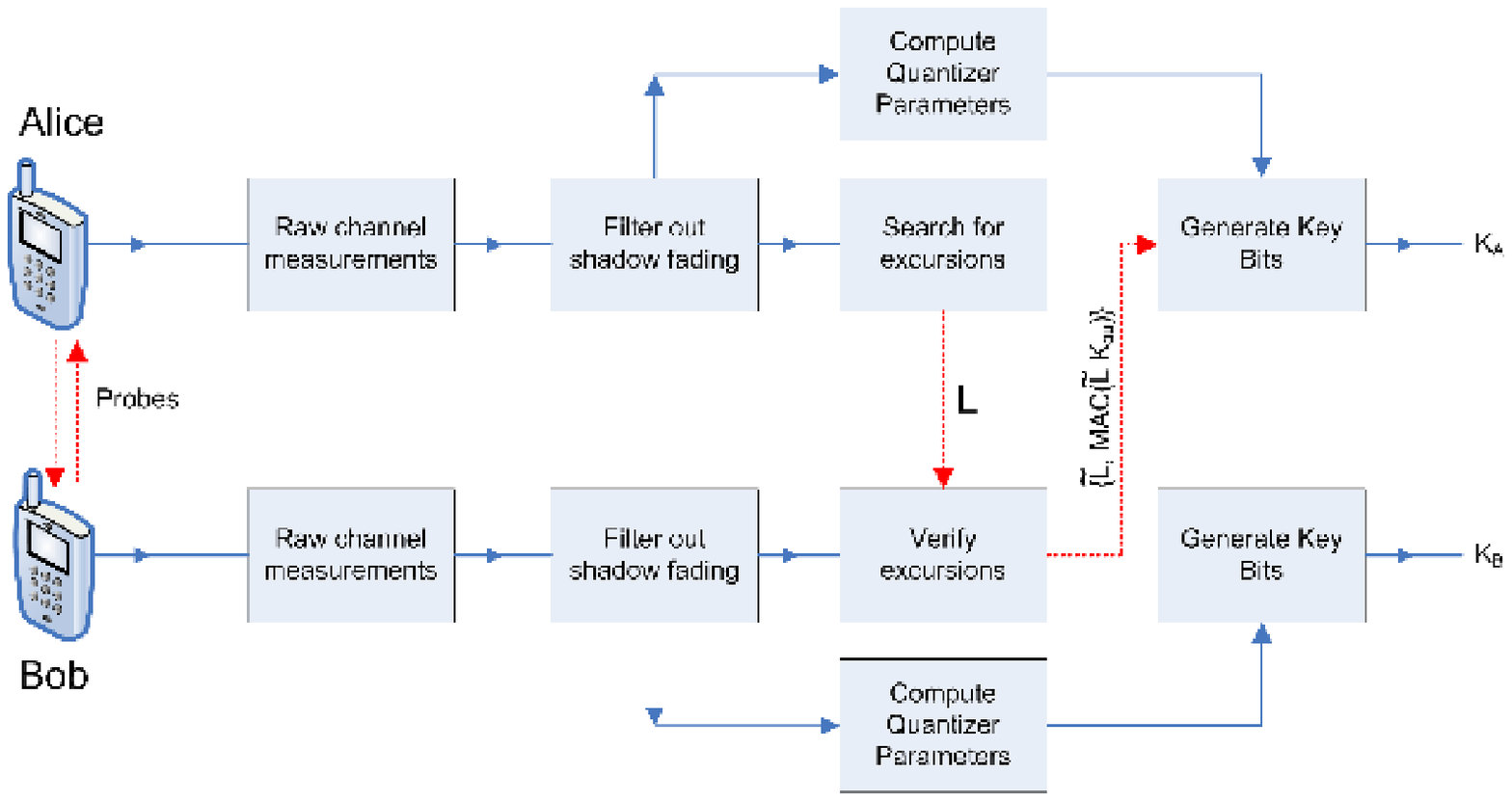}
\caption{A system level description of the level crossing algorithm.
Messages exchanged over the air are shown in dotted
lines.}\label{LC_sys}
 \end{center}
\end{figure}

Figure \ref{LC_sys} shows the system-level operation of the level
crossing algorithm. We show later that provided the levels $q_+,
q_-$ and the parameter $m$ are properly chosen, the bits generated
by the two users are identical with very high probability. In this
case, both Alice and Bob are able to compute identical key bits and
identical authentication key bits $K_{au}$, thereby allowing Alice
to verify that the protocol message $\tilde{L}$ did indeed come from
Bob.  Since Eve's observations from the channel probing do not
provide her with any useful information about $X^n$ and $Y^n$, the
messages $L$ and $\tilde{L}$ do not provide her any useful
information either. This is because they contain time indices only,
whereas the generated bits depend upon the values of the channel
estimates at those indices.


\subsection{Security Discussion for the Level-crossing Algorithm}
The secrecy of our key establishment method is based on the assumption that Alice and Bob have confidence that there is no eavesdropper Eve located near either Alice or Bob. Or equivalently, any eavesdropper is located a sufficient distance away from both Alice and Bob. In particular, the fading process associated with a wireless channel in a richly scattering environment decorrelates rapidly with distance and, for two receivers located at a distance of roughly the carrier wavelength from each other, the fading processes they each witness with respect to a transmitter will be nearly independent of each other\cite{Jak74}. For a Rayleigh fading channel model, if $h_{ba}$ and $h_{be}$ are the jointly Gaussian channels observed by Alice and Eve due to a probe transmitted by Bob, then the correlation between $h_{ba}$ and $h_{be}$ can be expressed as a function of the distance $d$ between Alice and Eve, and is given by  $J_0({2\pi d/\lambda})$, where $J_0(x)$ is the zeroth-order Bessel function of the first kind, $d$ is the distance between Alice and Eve, and $\lambda$ is the carrier wavelength. Hence, because of the decay of $J_0(x)$ versus the argument $x$, if we are given any $\epsilon>0$, it is possible to find the minimum distance $d$ that Eve must be from both Alice and Bob such that the mutual information $I(h_{ba}; h_{be}) \leq \epsilon$.

Further, we note that the statistical uniformity of the bit sequences that are extracted by Alice and Bob using our level-crossing algorithm is based on the statistical uniformity of positive and negative excursions in the distribution of the common stochastic channel between them. This inherently requires that the channel state representation for the fading process be symmetrically distributed about the distribution's mean. Many well-accepted fading models satisfy this property. Notably, Rayleigh and Rician fading channels\cite{AndreaGS}, which result from the multiple paths in a rich scattering environment adding up at the receiver with random phases, fall into this category. Consequently, we believe that the reliance of level-crossing algorithm on the underlying distribution symmetry, suggests that the level-crossing algorithm is best suited for Rayleigh or Rician fading environments. The independence of successive extracted bits follows from the fact that the excursions used for each bit are naturally separated by a coherence time interval or more, allowing the channel to decorrelate in time. Finally, we note that our approach does not preclude a final privacy amplification step, though application of such a post-processing step is straightforward and might be desirable in order to ensure that no information is gleaned by an eavesdropper.

\subsection{Performance Evaluation and Experimental Validation}

The central quantities of interest in our protocol are the rate of
generation of secret bits and the probability of error. The controls
available to us are the parameters: $q_+^u, q_-^u, m$ and the rate
at which Alice and Bob probe the channel between themselves, $f_s$.
We assume the channel is not under our control and the rate at which
the channel varies can be represented by the maximum Doppler
frequency, $f_d$. The typical Doppler frequency for indoor wireless
environments at the carrier frequency of $2.4$ GHz is $f_d =
\frac{v}{\lambda} \sim \frac{2.4 \times 10^9}{{3 \times 10^8}} = 8
\hspace{0.1cm}$ Hz, assuming a velocity $v$ of $1$ m/s. We thus
expect typical Doppler frequencies in indoor environments in the
$2.4$ GHz range to be roughly $10$ Hz. For automobile scenarios, we
can expect a Doppler of $\sim 200$ Hz in the $2.4$ GHz range. We
assume, for the sake of discussion, that the parameter of interest,
$F$ is a Gaussian random variable and the underlying stochastic
process $F(t)$ is a stationary Gaussian process. A Gaussian
distribution for $F$ may be obtained, for example, by taking $F$ to
be the magnitude of the in-phase component of a Rayleigh fading
process between Alice and Bob \cite{Rap01}. We note that the
assumption of a Gaussian distribution on $F$ is for ease of
discussion and performance analysis, and our algorithm is
valid in the general case where the distribution is symmetric about the mean.

The probability of error, $p_e$ is critical to our protocol. In
order to achieve a robust key-mismatch probability $p_k$, the
bit-error probability $p_e$ must be much lower than $p_k$. A
bit-error probability of $p_e = 10^{-7} \sim 10^{-8}$ is desirable
for keys of length $N=128$ bits. The probability of bit-error, $p_e$
is the probability that a single bit generated by Alice and Bob is
different at the two users. Consider the probability that the
$i^{th}$ bit generated  by Bob is ``$K_B^i = 0$'' at some index
given that Alice has chosen this index, but she has generated the
bit ``$K_A^i = 1$''. As per our Gaussian assumption on the parameter
$F$ and estimates $X$ and $Y$, this probability can be expanded as
\begin{eqnarray}\label{dejavu2}
&&\Pr(K_B^i = 0 | K_A^i = 1) =  \frac{\Pr(K_B^i=0, K_A^i =
1)}{\Pr(K_A^i = 1)} = \\ \nonumber && \frac{
\underbrace{\int_{q_+^X}^{\infty}\int_{-\infty}^{q_{-}^Y} \ldots
\int_{q_+^X}^{\infty}}_{(2m-1)~terms} \frac{(2
\pi)^{{(1-2m)/2}}}{|K_{2m-1}|^{1/2}}\exp{\left\{-\frac{1}{2}x^TK_{2m-1}^{-1}x
\right\}} d^{(2m-1)}x}{ \underbrace{\int_{q_+^X}^{\infty}\ldots
\int_{q_+^X}^{\infty}}_{(m)~terms} \frac{(2 \pi)^{-m/2}}{
|K_{m}|^{1/2}}\exp{\left\{-\frac{1}{2}x^TK_{m}^{-1}x \right\}}
d^{(m)}x},
\end{eqnarray}
where $K_m$ is the covariance matrix of $m$ successive Gaussian
channel estimates of Alice and $K_{2m-1}$ is the covariance matrix
of the Gaussian vector $(X_1, Y_1, X_2, \ldots, Y_{m-1}, X_{m})$
formed by combining the $m$ channel estimates of Alice and the $m-1$
estimates of Bob in chronological order. The numerator in
(\ref{dejavu2}) is the probability that of $2m-1$ successive channel
estimates ($m$ belonging to Alice, and $m-1$  for Bob), all $m$ of
Alice's estimates lie in an excursion above $q_{+}$ while all $m-1$
of Bob's estimates lie in an excursion below $q_{-}$. The
denominator is simply the probability that all of Alice's $m$
estimates lie in an excursion above $q_+$.

\begin{figure}
\begin{center}
 \includegraphics[scale  = 0.45]{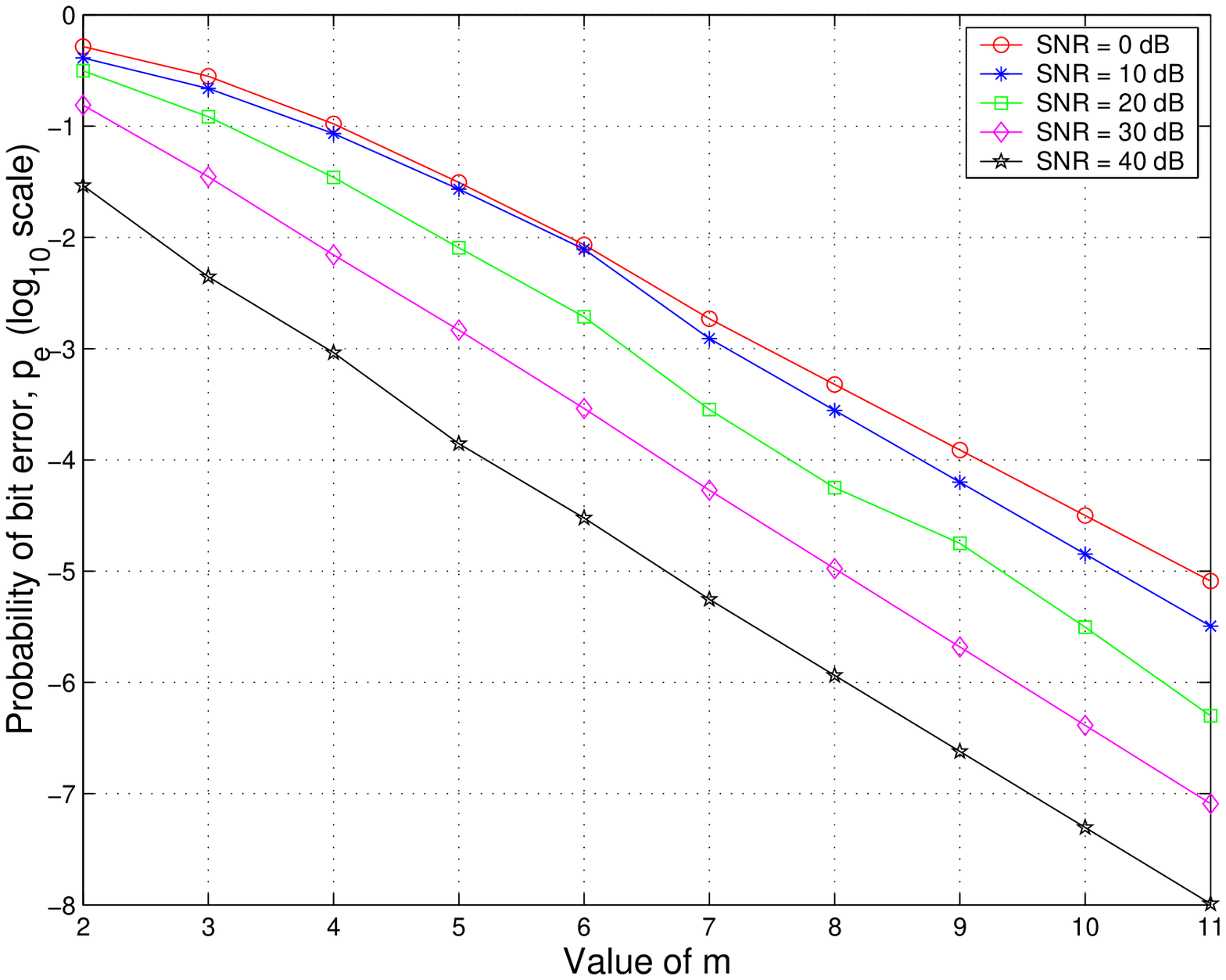}
\vspace{-0.3cm} \caption{Probability of bit error $p_e$ for various
values of m at different SNR levels ($\alpha = 0.8$ in
(\ref{q_def_u}), (\ref{q_def_u2}))}\label{pe2} \vspace{-0.45cm}
\end{center}
\end{figure}

We compute these probabilities for various values of $m$ and present
the results of the probability of error computations in Figure
\ref{pe2}. The results confirm that a larger value of $m$ will
result in a lower probability of error, as a larger $m$ makes it
less likely that Alice's and Bob's estimates lie in opposite types
of excursions. Note that if either user's estimates do not lie in an
excursion at a given index, a bit error is avoided because that
index is discarded by both users.

\begin{figure}
\begin{center}
 \includegraphics[scale = 0.5]{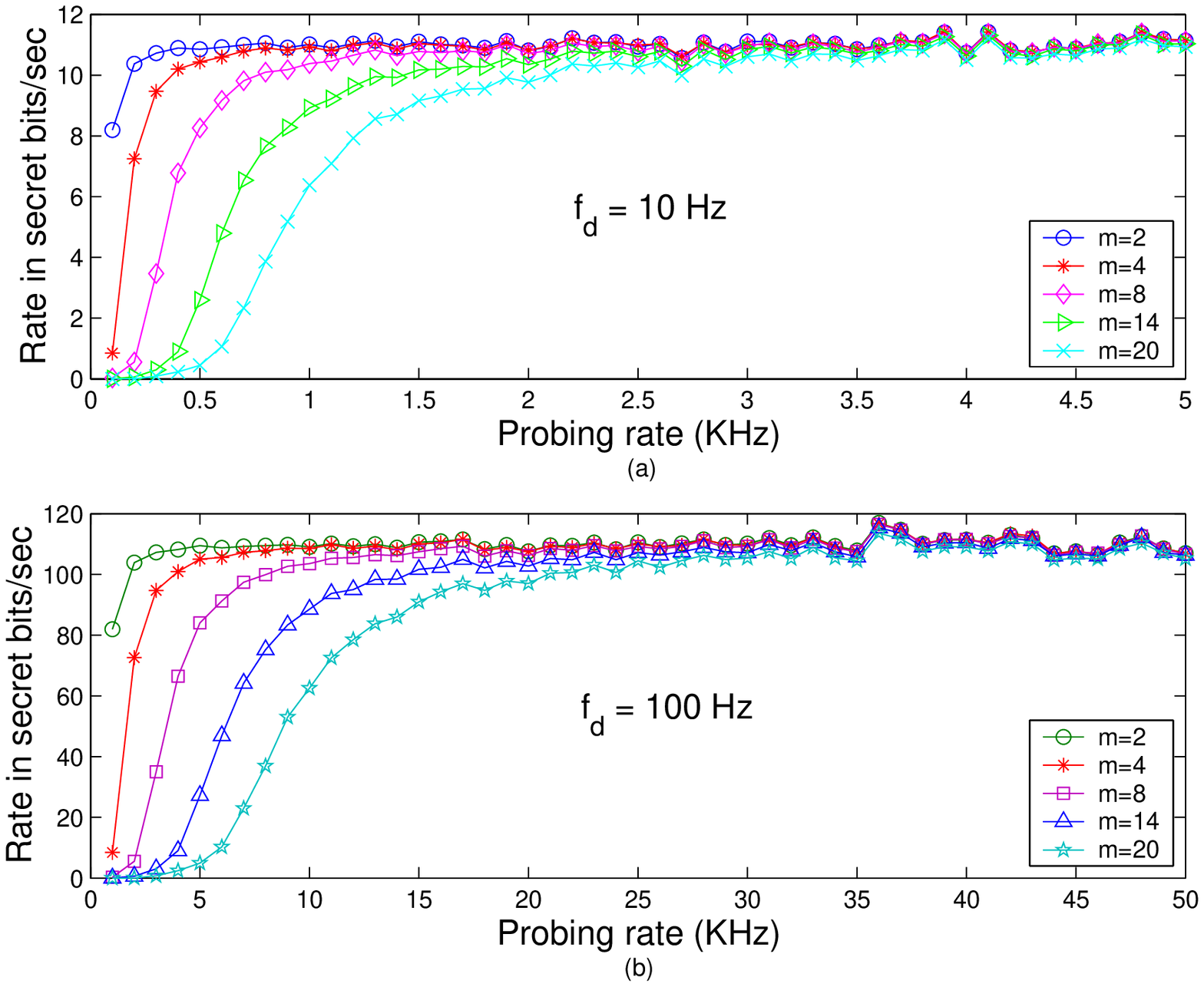}
\vspace{-0.27cm} \caption{Rate in secret bits per second for various
values of $m$, against probing rate for a channel with Doppler
frequency (a) $f_d = 10$ Hz and (b) $f_d = 100$ Hz ($\alpha = 0.8$
in (\ref{q_def_u}), (\ref{q_def_u2})).}\label{rate1}
\end{center}
\vspace{-0.75cm}
\end{figure}

How many secret bits/second (bps) can we expect to derive from a
fading channel using level crossings? An approximate analysis can be
done using the level-crossing rate for a Rayleigh fading process,
given by $LCR = \sqrt{2\pi}f_d \rho e^{- \rho^2}$ \cite{Rap01},
where $f_d$ is the maximum Doppler frequency and $\rho$ is the
threshold level, normalized to the root mean square signal level.
Setting $\rho =1$, gives $LCR \sim f_d$. This tells us that we
cannot expect to obtain more secret bits per second than the order
of $f_d$. In Figure \ref{rate1} (a) and (b), we plot the rate in
s-bits/sec as a function of the channel probing rate for a Rayleigh
fading channel with maximum Doppler frequencies of $f_d = 10$ Hz and
$f_d = 100$ Hz respectively. As expected, the number of s-bits the
channel yields increases with the probing rate, but saturates at a
value on the order of $f_d$.

In order for successive bits to be statistically independent, they
must be separated in time by more than one coherence time interval.
While the precise relationship between coherence time and Doppler
frequency is only empirical, they are inversely related and it is
generally agreed that the coherence time is smaller in magnitude
(Coherence time $T_c$, is sometimes expressed in terms of $f_d$ as
$T_c \thickapprox \sqrt{\frac{9}{16 \pi f_d^2}}$) than $1/f_d$.
Therefore, on average, if successive bits are separated by a time
interval of $1/f_d$, then they should be statistically independent.

More precisely, the number of secret bps is the number of secret
bits per observation times the probing rate. Therefore
\begin{eqnarray}
R_k & \hspace{-0.25cm}= &\hspace{-0.2cm} H(bins) \times \Pr(K_A^i=K_B^i) \times \frac{f_s}{m} \\ \nonumber
&\hspace{-0.25cm} = &\hspace{-0.2cm} 2 \frac{f_s}{m} \times
\Pr(K_A^i = 1, K_B^i=1) \nonumber \\ &\hspace{-0.25cm}=
&\hspace{-0.3cm} 2\frac{f_s}{m}.
\hspace{-0.15cm}\underbrace{\int_{q_+^X}^{\infty}\hspace{-0.23cm}
\ldots \int_{q_+^X}^{\infty}}_{(2m-1)~terms} \hspace{-0.1cm}\frac{(2
\pi)^{\frac{1-2m}{2}}}{|K_{2m-1}|^{1/2}}e^{\left\{-\frac{1}{2}x^TK_{2m-1}^{-1}x
\right\}} d^{2m-1}x ,\nonumber
\end{eqnarray}
where $H(bins)$ is the entropy of the random variable that
determines which bin ($>q_+$ or $<q_-$) of the quantizer the
observation lies in, which in our case equals $1$ assuming that the
two bins are equally likely\footnote{The levels $q_+$ and $q_-$ are
chosen so as to maintain equal probabilities for the two bins.}. The
probing rate $f_s$ is normalized by a factor of $m$ because a single
`observation' in our algorithm is a sequence of $m$ channel
estimates.

\begin{figure}
\begin{center}
\begin{tabular}[c]{p{6 cm}p{6cm}}
\begin{center}
 \includegraphics[scale=0.42]{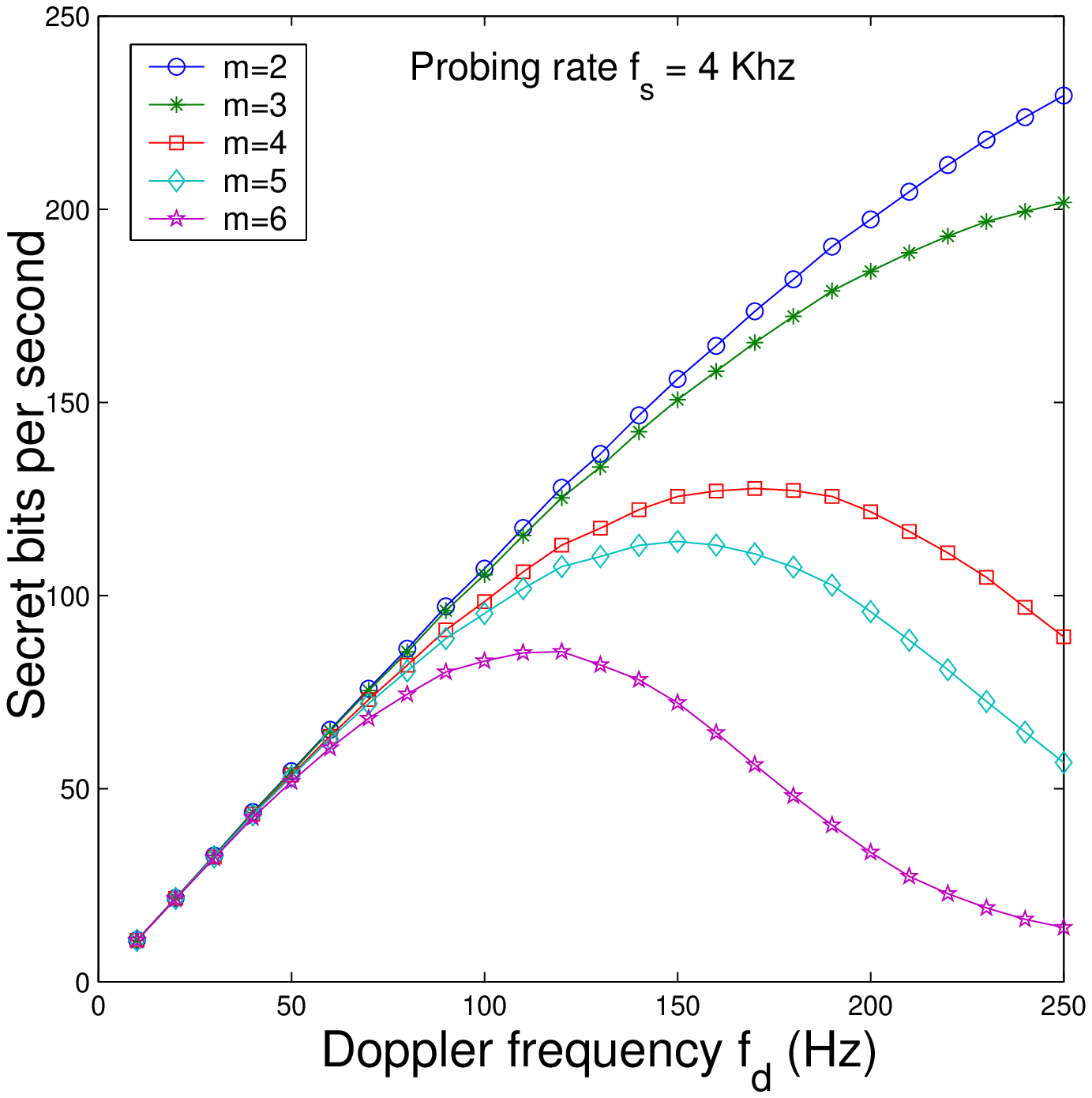}
\\ \scriptsize{(a)}
\end{center}
&
\begin{center}
\includegraphics[scale=0.42]{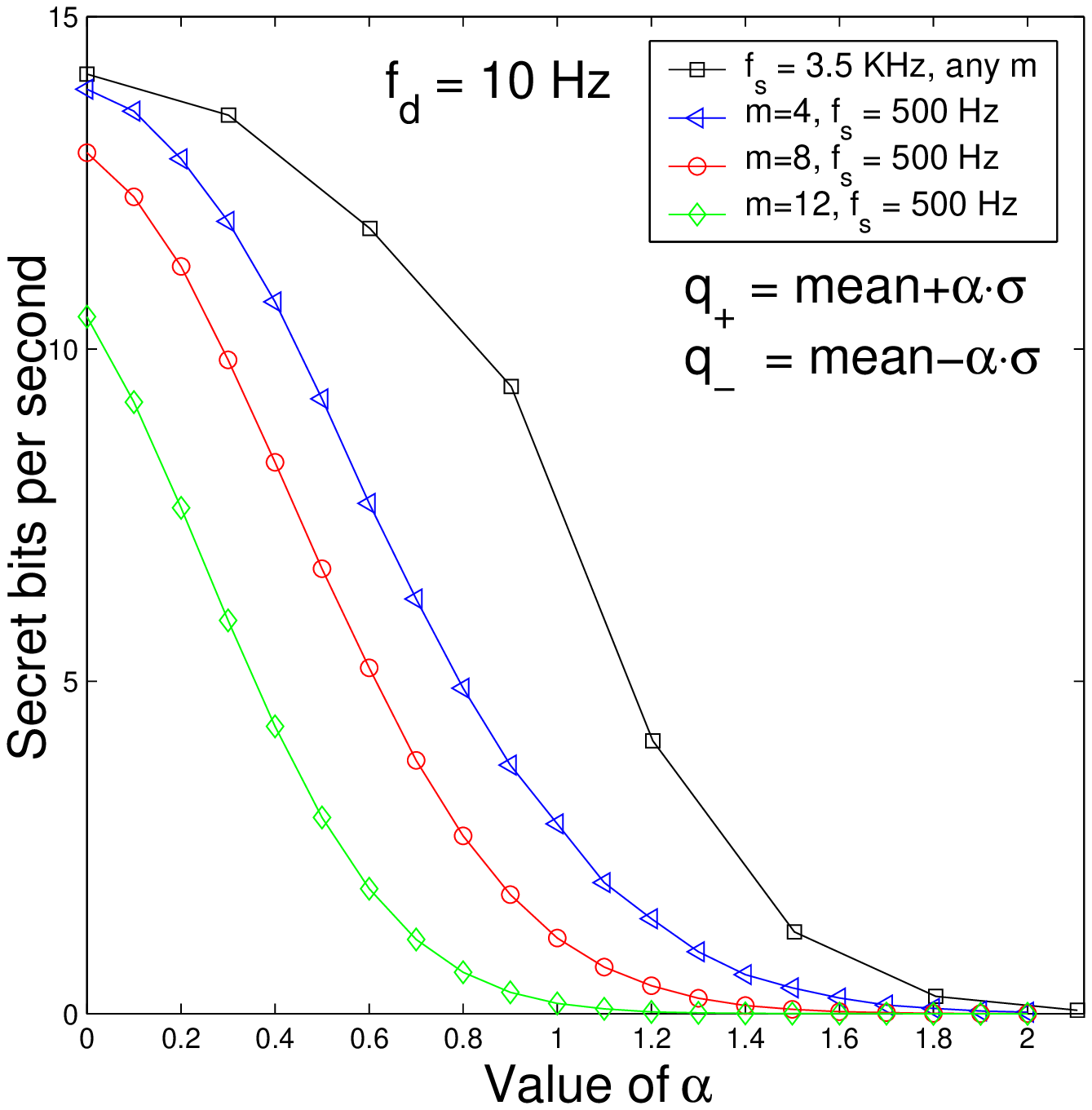}
\\ \scriptsize{(b)}
\end{center}
\end{tabular}
\end{center}
\vspace{-0.6cm}
 \caption{(a) Secret-bit rate for varying
Doppler $f_d$ and fixed $f_s$ for various values of $m$ (b) Rate as
a function of function of quantizer levels $q_+$ \& $q_-$
parametrized by $\alpha$.} \vspace{-0.25cm} \label{rate6}
\end{figure}

Figure \ref{rate1} confirms the intuition that the secret bit rate
must fall with increasing $m$, since the longer duration excursions
required by a larger value of $m$ are less frequent. In Figure
\ref{rate6}(a), we investigate how the secret-bit rate $R_k$ varies
with the maximum Doppler frequency $f_d$, i.e., the channel
time-variation. We found that for a fixed channel probing rate (in
this case, $f_s = 4000$ probes/sec), increasing $f_d$ results in a
greater rate but only up to a point, after which the secret-bit rate
begins to fall. Thus, `running faster' does not necessarily help
unless we can increase the probing rate $f_s$ proportionally. Figure
\ref{rate6}(b) shows the expected decrease in secret-bit rate as the
quantizer levels the value of $\alpha$ is varied to move $q_+^u$ and
$q_-^u$ further apart. Here, $\alpha$ denotes the number of standard
deviations from the mean at which the quantizer levels are placed.

We examined the performance of the secrecy generation system through
experiments. The experiments involved three terminals, Alice, Bob
and Eve, each equipped with an 802.11a development board.

In the experiments, Alice was configured to be an access point (AP),
and Bob was configured to be a station (STA). Bob sends Probe
Request messages to Alice, who replies with Probe Response messages
as quickly as possible. Both terminals used the long preamble
segment \cite{80211a} of their received Probe Request or Probe
Response messages to compute 64-point CIRs. The tallest peak in each CIR (the dominant multipath) was used as the
channel parameter of interest, i.e., the $X$ and $Y$ sample inputs
to the secret key generation system. To access such peak data,
FPGA-based customized logic was added to the 802.11 development
platform. Eve was configured to capture the Probe Response messages
sent from Alice in the experiments.

Two experiments were conducted. In the first experiment, Alice and
Eve were placed in a laboratory. In a second experiment, Alice and
Eve remained in the same positions while Bob circled the cubicle
area of the office.

\begin{figure}
          \centering
          \subfigure[Alice, Bob and Eve's $64$-point CIRs from a
          common pair of Probe Request, Probe Response messages.]{
          \includegraphics[width=5in, height=1.3in]{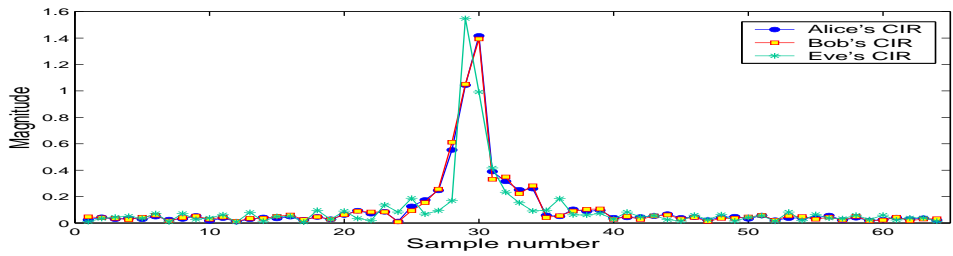}}

          \subfigure[Traces of the magnitudes resulting from 200 Alice, Bob and Eve's CIRs.]{
          \includegraphics[width=6.8in, height=1.5in]{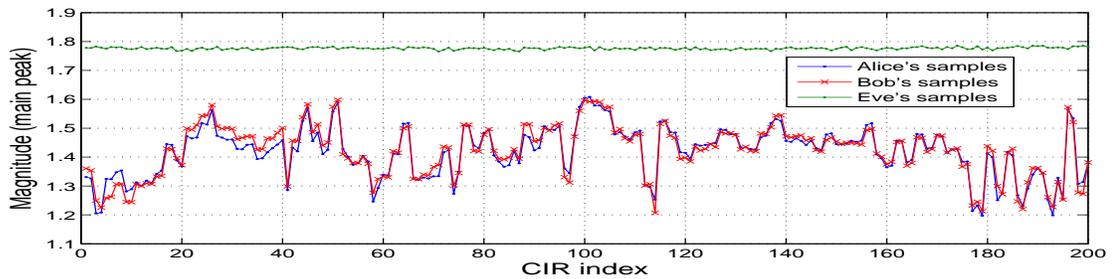}}
           \caption{Some examples of experimental data.}
           \label{ABECIR}
         \end{figure}

Figure \ref{ABECIR}(a) shows an example of Alice's, Bob's, and Eve's
64-point CIRs obtained through a single common pair of Probe Request and
Probe Response messages. It is seen from the figure that Alice's and
Bob's CIRs look similar, while they both look different from Eve's
CIR. We show the traces for Alice and Bob resulting from 200 consecutive CIRs in Figure \ref{ABECIR}(b). The
similarity of Alice's and Bob's samples, as well as their difference
from Eve's samples, are evident from the figure.

While our experiments ran for $\sim22$ minutes, in the interest of
space and clarity we show only $700$ CIRs collected over a duration of
$\sim77$ seconds. Each user locally computes $q_{+}$ and $q_{-}$ as
in (\ref{q_def_u}), (\ref{q_def_u2}). We chose $\alpha =
\frac{1}{8}$ for our experiments.

\begin{figure}
 \begin{center}
\includegraphics[width=5in, height=3.2in]{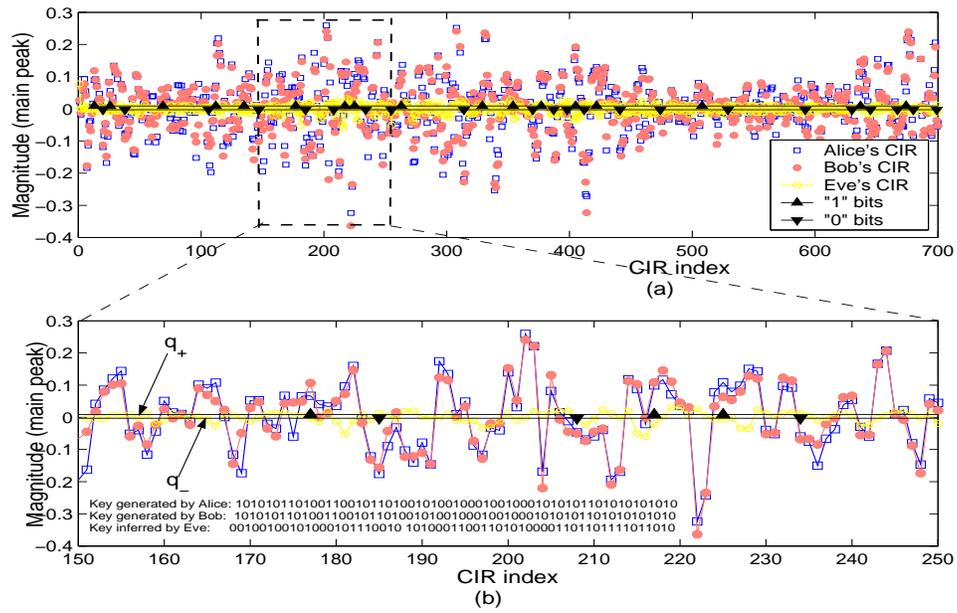}
\caption{\label{IDs2}(a) Traces of Alice and Bob after subtracting
average signal power. Using $m=5$, $N=59$ bits were generated in
$110$ seconds ($R_k = 0.54$ s-bits/sec) while $m=4$ gives $N=125$
bits ($R_k = 1.13$ s-bits/sec.) with no errors in each case. (b) A
magnified portion of (a)}
\end{center}
\end{figure}

Figure \ref{IDs2} shows the traces collected by Alice and Bob after
removal of slow shadow fading components using a simple local
windowed mean. This is to prevent long strings of $1$s and $0$s, and
to prevent the predictable component of the average signal power
from affecting our key generation process. Using the small scale
fading traces, our algorithm generates $N=125$ bits in $110$ seconds
($m=4$), yielding a key rate of about $1.13$ bps. Figure \ref{IDs2}
shows the bits that Eve would generate if she carried through with
the key-generation procedure. The results from our second experiment with a moving Bob are very
similar to the ones shown for the first experiment, producing $1.17$
bps. with $m=4$ and $\alpha = \frac{1}{8}$. Note that while figures \ref{rate1} and \ref{rate6} depict the secret bit rate that can be achieved for the specified values of Doppler frequency, our experimental setup does not allow us to measureably control the precise Doppler frequency and the secret bits rates we report from our experiments correspond only the indoor channel described.

In order to verify the assumption that Eve does not gain any useful
information by passive observation of the probes transmitted by
Alice and Bob, we empirically computed the mutual information using the
method in \cite{WanKul06} between the signals received at the
legitimate users and compare this with that between the signals
received by Eve and a legitimate user. The results of this
computation, summarized in Table \ref{TableMI}, serve as an upper
bound to confirm that Eve does not gather any significant
information about the signals received at Alice and Bob. Although this information leakage is minimal relative to the mutual information shared between Alice and Bob, it might nonetheless be prudent to employ privacy amplification as a post-processing to have a stronger assurance that Eve has learned no information about the key established between Alice and Bob. Finally, we
note that with suitable values of the parameters chosen for the
level crossing algorithm, the bits extracted by Alice and Bob are
statistically random and have high-entropy per bit. This has been
tested for and previously reported in \cite{MatTra08} using a suite
of statistical randomness tests provided by NIST \cite{NIST}.

\section{Quantization-Based Secret Key Generation for Wireless Channels}

We now present a more powerful and general approach than the level-crossing approach
discussed in Section II for obtaining secret keys
from the underlying fading phenomena associated with a with a richly scattering wireless environment. Whereas the level-crossing algorithm was best suited for extracting keys from channel states whose distributions are inherently symmetric, our second approach is applicable to more general channel state distributions. Further, this second approach approach is capable of
generating significantly more than a single bit per independent
channel realization, especially when the channel estimation SNRs are
high.

To accomplish this we propose a new approach for the quantization of
sources whose statistics are not known, but are believed to be
similar in the sense of having ``high SNR"  - a notion we shall
define more precisely below.  Our quantization approach is motivated
by considering a simpler setting of a Gaussian source model and
addressing certain deficiencies which can be observed in that model.
This problem has been addressed by \cite{YeRez06} using a simple
``BICM-like" approach \cite{BloTha06} to the problem. A more general
treatment which introduces multi-level coding can be found in
\cite{BloTha06} and also \cite{BloBar08}, however for our purposes,
the simple "BICM-like" approach of \cite{YeRez06} and
\cite{BloTha06} is sufficient.  To motivate our approach to
``universal'' quantization we need to take this solution and improve
on it - the process which we describe next.

\subsection{Over-quantized Gaussian Key Generation System}

\begin{figure}[h]
          \centering
          \subfigure[Alice's secrecy processing.]{
          \includegraphics[width=5.5in, height=2.2in]{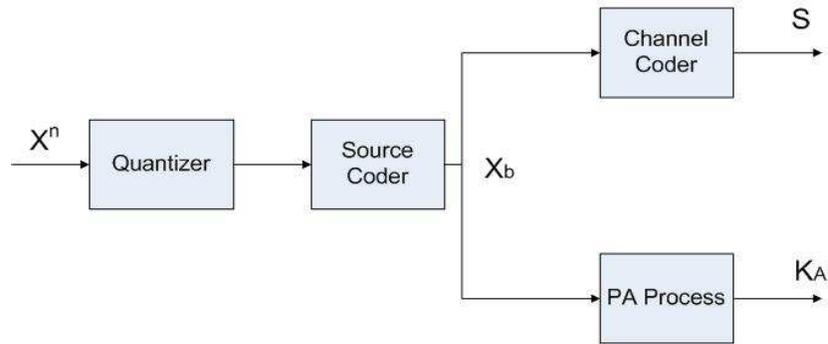}}

          \subfigure[Bob's secrecy processing.]{
          \includegraphics[width=5in, height=0.8in]{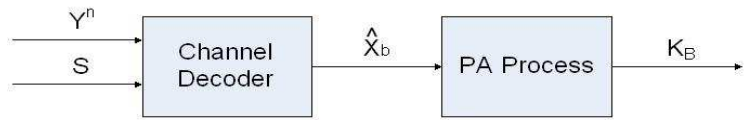}}
           \caption{Block diagrams of the basic system.}
           \label{fig1}
         \end{figure}

We begin our discussion of the over-quantized Gaussian Key
Generation System by reviewing the simple approach to the problem
described in \cite{YeRez06}. A block diagram of a basic secret key
generation system is shown in Figure \ref{fig1}. Alice's secrecy
processing consists of four blocks: Quantizer, Source Coder, Channel
Coder and the Privacy Amplification (PA) process. The Quantizer
quantizes Alice's Gaussian samples $X^n$. The Source Coder converts
the quantized samples to a bit string ${\bf X}_b$. The Channel Coder
computes the syndrome ${\bf S}$ of the bit string ${\bf X}_b$.  A
rate $1/2$ LDPC code is used in \cite{YeRez06}. This syndrome is
sent to Bob for his decoding of ${\bf X}_b$. As discussed in Section
I, the transmission of the syndrome is assumed to take place through
an error-free public channel; in practice this can be accomplished
through the wireless channel with the use of standard reliability
techniques (e.g., CRC error control and ARQ). Finally, privacy
amplification (if needed) is implemented in the PA block.

\begin{figure}
         \begin{center}
         \begin{tabular}[c]{p{7 cm}p{7.3cm}}
         \begin{center}
          \includegraphics[width = 3.2in, height=2.8in]{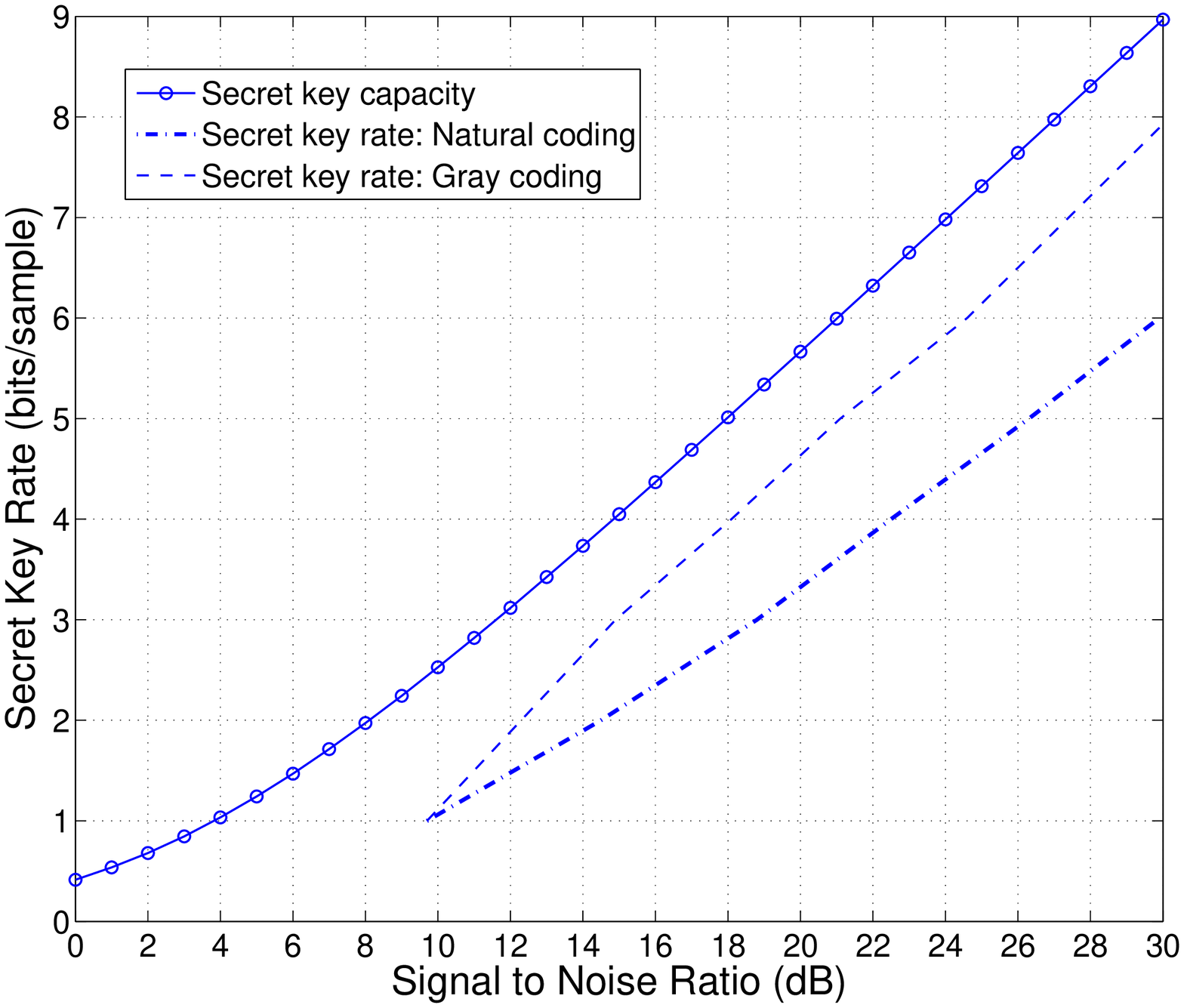}
          \\ \scriptsize{(a)}
          \end{center}
          &
          \begin{center}
            \includegraphics[width=3.2in, height=2.8in]{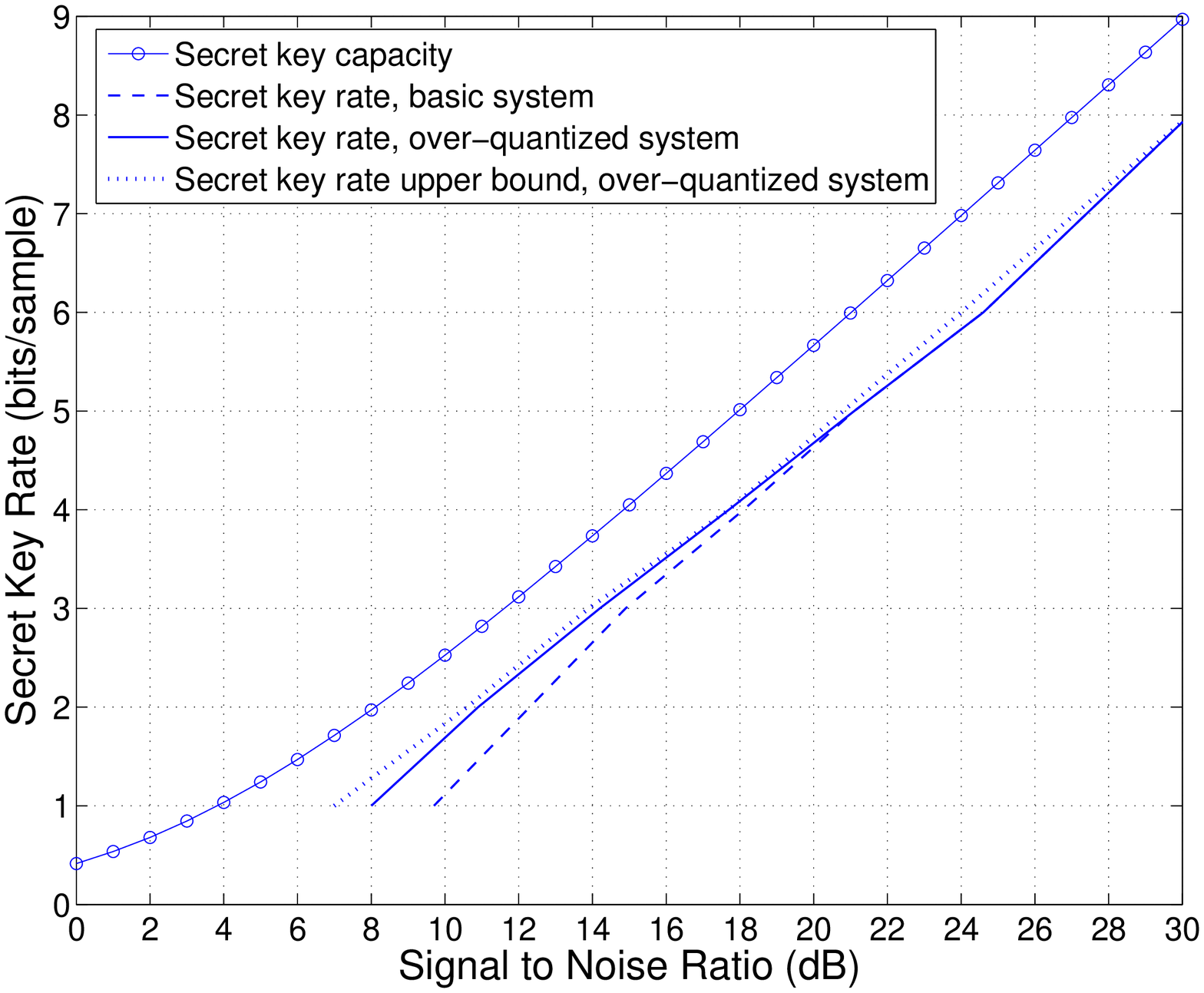}
            \\ \scriptsize{(b)}
            \end{center}
            \end{tabular}
            \end{center}
            \vspace{-0.75cm} \caption{\label{KeyratesGaussian}(a) Secret key rates achieved
            by the basic system. (b) Secret key rates achieved by the improved system.}
            \end{figure}

Figure \ref{KeyratesGaussian}(a) present the results obtained by
using various algorithm options discussed in \cite{YeRez06}. We
observe from this figure that at high SNR ($>15 $dB), the secret key
rates resulting from Gray coding are within 1.1 bits of the secret
key capacity (\ref{e2.6}). However, the gap between the achieved
secret key rates and the secret key capacity is larger at low SNR.
In this sub-section, we demonstrate how the basic system can be
improved such that the gap at low SNR is reduced. We restrict
ourselves to Gray coding, as this is clearly the better source
coding approach.

We start with the observation that the quantization performed by
Alice involves some information loss. To compensate for this, Alice
could quantize her samples at a higher level than the one apparently
required for the basic secret key generation purpose. Suppose that
quantization to $v$ bits is required by the baseline secrecy
generation scheme. Alice then quantizes to $v+m$ bits using Gray
coding as a source coder.  We refer to the $v$ most significant bits
as the \emph{regularly quantized bits} and the $m$ least significant
bits as the \emph{over-quantized bits}. The over-quantized bits
${\bf B}$ are sent directly to Bob through the error-free public
channel.

The Channel Decoder (at Bob) uses the syndrome ${\bf S}$ of the
regularly quantized bits ${\bf X}_b$, the over-quantized bits ${\bf
B}$ and Bob's Gaussian samples $Y^n$ to decode ${\bf X}_b$. Again,
it applies the modified belief-propagation algorithm (cf.
\cite{LivXio02}), which requires the per-bit LLR. The LLR
calculation is based on both $Y^n$ and ${\bf B}$.

Suppose one of Alice's Gaussian samples $X$ is quantized and Gray
coded to bits $(X_{b,1}, \cdots, X_{b, v+m})$. With Bob's
corresponding Gaussian sample $Y$ and Alice's over-quantized bits
$(X_{b,v+1}, \cdots, X_{b, v+m}) = (a_{v+1}, \cdots, a_{v+m})$, the
probability of $X_{b,i}$, $1\leq i\leq v$, being 0 is derived below:
\begin{eqnarray}
&&\Pr(X_{b,i}=0|Y=y,X_{b,v+1}=a_{v+1}, \cdots, X_{b,v+m} = a_{v+m})\nonumber\\
& = & \frac{\Pr(X_{b,i}= 0, X_{b, v+1}= a_{v+1},\cdots, X_{b,
v_m}=a_{v+m}| Y=y)}
{\Pr(X_{b, v+1}= a_{v+1},\cdots, X_{b, v_m}=a_{v+m}| Y=y)}\nonumber\\
&=& \frac{\sum_{j=1}^{2^{v+m}}\Pr(\bar{q}_{j-1}\leq X<
\bar{q}_j|Y=y){\bf 1}_{G_{v+m}^{i}(j-1)=0} \cdot {\bf
1}_{G_{v+m}^{v+1}(j-1)=a_{v+1}} \cdots {\bf
1}_{G_{v+m}^{v+m}(j-1)=a_{v+m}}}
{\sum_{j=1}^{2^{v+m}}\Pr(\bar{q}_{j-1}\leq X< \bar{q}_j|Y=y)\cdot
{\bf 1}_{G_{v+m}^{v+1}(j-1)=a_{v+1}} \cdots {\bf
1}_{G_{v+m}^{v+m}(j-1)=a_{v+m}}}, \label{e3.8}
\end{eqnarray}
where ${\bf 1}$ is an indicator function and the function
$G_{k}^{i}(j)$, $1\leq i\leq k$, $0\leq j\leq 2^k-1$, denotes the
$i^{th}$ bit of the $k$-bit Gray codeword representing the integer
$j$. The quantization boundaries $\bar{q}_0 < \cdots
<\bar{q}_{2^{v+m}}$ depend on the quantization scheme used. For
instance, the quantization boundaries of the {\it equiprobable
quantizer} satisfy
\begin{equation}
\int_{\bar{q}_{j-1}}^{\bar{q}_j} \frac{1}{{\sqrt {2\pi N}
}}e^{-\frac{x^2}{2N}}dx = \frac{1}{2^{v+m}}, \ \
 j= 1, \cdots , 2^{v+m}.
\end{equation}
Now,
\begin{eqnarray}
\Pr(\bar{q}_{j-1}\leq X< \bar{q}_j|Y=y)
&=& \Pr(\bar{q}_{j-1}\leq \frac{P}{P+N}Y+Z_0<\bar{q}_j|Y=y)\nonumber\\
&=& \Pr(\bar{q}_{j-1}-\frac{P}{P+N}y\leq
Z_0<\bar{q}_j-\frac{P}{P+N}y) \nonumber\\
&=& Q\left(\frac{\bar{q}_{j-1}-\frac{P}{P+N}y}{{\sqrt
\frac{2PN+N^2}{P+N}}}\right)-
Q\left(\frac{\bar{q}_{j}-\frac{P}{P+N}y}{{\sqrt
\frac{2PN+N^2}{P+N}}}\right)\nonumber\\
&=& g(j-1, y)-g(j,y)\nonumber,
\end{eqnarray}
where the function $g(k,y)$, $0\leq k\leq 2^{v+m}$, is defined as
\begin{equation}
g(k,y)=Q\left(\frac{\bar{q}_k-\frac{P}{P+N}y}{\sqrt{(2PN+N^2)/(P+N)}}\right),
\end{equation}
and $Q$ is the usual Gaussian tail function \cite{Proakis00}. Hence,
the probability of (\ref{e3.8}) is given by
\begin{equation}
\frac {\sum_{j=1}^{2^{v+m}}\left[g(j-1,y)-g(j,y)\right]\cdot {\bf
1}_{G_{v+m}^{i}(j-1)=0} \cdot {\bf 1}_{G_{v+m}^{v+1}(j-1)=a_{v+1}}
\cdots {\bf 1}_{G_{v+m}^{v+m}(j-1)=a_{v+m}}}
{\sum_{j=1}^{2^{v+m}}\left[g(j-1,y)-g(j,y)\right]\cdot {\bf
1}_{G_{v+m}^{v+1}(j-1)=a_{v+1}} \cdots {\bf
1}_{G_{v+m}^{v+m}(j-1)=a_{v+m}}}. \label{LLR}
\end{equation}
It should be noted that when equiprobable quantization is used, the
over-quantized bits ${\bf B}$ and the regularly quantized bits ${\bf
X}_b$ are independent as shown below. Suppose a sample $X$ is
equiprobably quantized and source coded to $t$ bits $(X_{b,1}, \cdots
, X_{b,t})$. For an arbitrary bit sequence $(a_1, \cdots, a_{t})$
and a set ${\cal S}\subseteq {\cal T}= \{1,\cdots , t\}$, we have
\begin{eqnarray}
&&\Pr \left(\{X_{b,i} = a_i: i\in {\cal S}\}|\{X_{b,i} = a_i: i\in
{\cal T}\setminus{\cal S}\} \right)
= \frac{\Pr\left(\{X_{b,i} = a_i: i\in {\cal T}\}\right)}{\Pr\left(\{X_{b,i} = a_i: i\in {\cal T}\setminus{\cal S}\}\right)}\nonumber\\
&=&\frac{2^{-t}}{2^{-(t-|{\cal S}|)}}=2^{-|{\cal
S}|}=\Pr\left(\{X_{b,i} = a_i: i\in {\cal S}\}\right), \nonumber
\end{eqnarray}
which implies the amount of secrecy information remaining in ${\bf
X}_b$ after the public transmission is at least $|{\bf X}_b|- |{\bf
S}|$ bits.\footnote{Relying on hash functions for privacy
amplification requires the use of R\'{e}nyi entropy. However, we can
use \cite[Theorem 3]{BenBra95} to equivocate R\'{e}nyi and Shannon
entropies.} Note that this conclusion does not hold for other
quantization approaches (e.g., MMSE quantization) and, therefore,
equiprobable quantization should be used if over-quantization is
applied.

On the other hand, it is implied by (\ref{LLR}) that the
over-quantized bits ${\bf B}$ and the regularly quantized bits ${\bf
X}_b$ are dependent given Bob's samples $Y^n$. Hence, $I({\bf X}_b;
{\bf B}|Y^n)>0$. It follows from the Slepian-Wolf theorem (cf.
\cite{CovTho91}) that with the availability of the over-quantized
bits ${\bf B}$, the number of syndrome bits $|{\bf S}|$ required by
Bob to successfully decode ${\bf X}_b$ is approximately $H({\bf
X}_b|Y^n, {\bf B})$, which is less than $H({\bf X}_b|Y^n)$, the
number of syndrome bits transmitted in the basic system. In other
words, the secret key rate achieved by the over-quantized system is
approximated by $\frac{1}{n}I({\bf X}_b; Y^n, {\bf B})$, which is
larger than $\frac{1}{n}I({\bf X}_b; Y^n)$, the secret key rate
achieved by the basic system.

To obtain an upper limit on the performance improvement that
over-quantization may provide us, we can imagine sending the entire
(real-valued) quantization error as a side information. There are a
number of issues with this approach. Clearly, distortion-free
transmission of real-valued quantities is not practically feasible.
However, as we are looking for a bound, we can ignore this. More
importantly, the transmission of raw quantization errors may reveal
information about ${\bf X}_b$. For example, to equiprobably quantize
a zero mean, unit variance Gaussian random variable with 1 bit per
sample, the quantization intervals are $(-\infty, 0]$ and
$(0,\infty)$, with respective representative value -0.6745 and
0.6745. Suppose a sample $X$ is of value 2, then its quantization
error is $2-0.6745=1.3255$. This implies that $X$ must be in the
interval $(0, \infty)$, since otherwise, the quantization error does
not exceed 0.6745. Thereby, it is necessary to process the raw
quantization errors such that the processed quantization errors do
not contain any information about ${\bf X}_b$. For this purpose, it
is desirable to transform quantization errors to uniform
distribution. To do so, we first process an input sample $X$ with
the cumulative distribution function (CDF) of its distribution and
then quantize. The transformed quantization error is then given by
$E=\phi\left(X\right)-\phi\left(q(X)\right)$, where $\phi(x)$ is the
CDF for $X$ and $q(X)$ is the representative value of the interval
to which $X$ belongs. The quantization errors
$E^n=(E_1,\cdots,E_n)$, which are then uniformly distributed on
$\left[-2^{-(v+1)}, 2^{-(v+1)}\right]$, are sent to Bob through the
error-free public channel.

The rest of the process (encoding/decoding and PA) proceeds as
before. However, the LLR computation must be modified to use
probability density functions, rather than probabilities:
\begin{equation}
\ln \frac{\Pr(X_{b,i}=0|Y=y, E=e)}{\Pr(X_{b,i}=1|Y=y, E=e)} =
\sum_{j=1}^{2^v} (-1)^{{\bf 1}_{G_{v}^{i}(j-1)=0}}\cdot
h(e,j,y),\label{e3.12}
\end{equation}
where the function $G_{k}^{i}(j)$ is defined in (\ref{e3.8}) and the
function $h(e,j,y)$ is defined as
\[
h(e,j,y)=\frac{P+N}{2(2PN+N^2)} \left(\phi^{-1}
\left(e+\frac{j-0.5}{2^v}\right) -\frac{P}{P+N} y \right)^2,
\]
for $-2^{-(v+1)}\leq e\leq 2^{-(v+1)}$, $1\leq j\leq 2^v$, with the
function $\phi$ being the CDF for $X$. The derivation of
(\ref{e3.12}) is similar to that of (\ref{LLR}), which is omitted
here.

Figure \ref{KeyratesGaussian}(b) shows simulation results for 2-bit
over-quantization and the upper bound. We note, as expected, that
the overall gap to capacity has been reduced to about 1.1 dB at the
low-SNR.

\subsection{A Universal Secret Key Generation System}

In the previous sub-section we discussed secret key
generation for a jointly Gaussian model. The random variables $X$ and
$Y$ in the model are jointly Gaussian distributed and the
distribution parameter SNR is known at both terminals. However, in many practical conditions, the
correlated random variables at the two terminals may not be subject
to a jointly Gaussian distribution, and the distribution parameters
are usually unknown or estimated inaccurately. 

We address this problem by describing a method for LLR generation
and subsequent secrecy generation that makes very few assumptions
on the underlying distribution.  As we shall see this method is largely
based on the over-quantization idea we introduced above.

\subsubsection{System Description}

Compared to the basic system (Figure \ref{fig1}) developed for the
Gaussian model, the universal system includes two additional Data
Converter blocks (one at Alice; the other at Bob), and modified
Quantizer and Channel Decoder blocks. The inputs to Alice's Data
Converter blocks are $X^n$ and the outputs of Alice's Data Converter
block are sent to the modified Quantizer block. The inputs to Bob's
Data Converter blocks are $Y^n$ and the outputs of Bob's Data
Converter block are sent to the modified Channel Decoder block.

The purpose of the Data Converter is to convert the input samples
$X^n$, $Y^n$ to uniformly distributed samples $U^n$, $V^n$, where
$U_i, V_i\in [0,1)$. The conversion is based on the empirical
distribution of input samples. Given the $i^{th}$ sample $X_i$ of
input samples $X^n$, denote by $K_n(X_i)$ the number of samples in
$X^n$ which are strictly less than $X_i$ plus the number of samples
in $X^n$ which are equal to $X_i$ but their indices are less than
$i$.  The output of the Data conversion block corresponding to $X_i$
is given by $U_i=\frac{K_n(X_i)}{n}$.

To justify the use of this approach, we show that $U^n$
asymptotically tends to an i.i.d. sequence, each uniformly
distributed between 0 and 1. Thus, while for any finite block length
the sequence $U^n$ is not comprised of independent variables, it is
assymtotically i.i.d. uniform.  Consider an i.i.d. sequence
$X^n=(X_1, \cdots ,X_n)$ . Denote by $\phi$ the actual CDF of $X_i$.
Let $W_i=\phi(X_i)$, $i=1,\cdots, n$. Then $W_1, \cdots, W_n$ is an
i.i.d. sequence, each uniformly distributed between 0 and 1. Hence,
it suffices to show that the sequence $U^n$ converges to the
sequence $W^n$.

Convergence of the empirical distribution to the true distribution
is a well-established fact in probability known as the
Glivenko-Cantelli Theorem \cite{Shorak86}.  However, we need a
stronger statement which gives the rate of such convergence. This is
known as the Dvoretzky-Kiefer-Wolfowitz Theorem \cite{DKW56} and is
stated in the following lemma.

\begin{lemma} \cite{DKW56} Let $X_1,\cdots, X_n$ be real-valued,
i.i.d. random variables with distribution function $F$. Let $F_n$
denote the associate empirical distribution function defined by
\[
F_n(x) = \frac{1}{n} \sum_{i=1}^n 1_{(-\infty, x]}(X_i),\ \  x\in
{\cal R}.
\]
For any $\varepsilon>0$,
\begin{equation}
\Pr\left(\sup_{x\in {\cal R}}|F_n(x)-F(x)|>\varepsilon \right)\leq
2e^{-2n\varepsilon^2}. \label{DKW}
\end{equation}
$\hfill \Box$
\label{LemmaDKW}
\end{lemma}

We will also need the notion of a $L^p$ convergence of random
sequences \cite{Chung01}. The $L^p$-norm of a sequence $X^n$, $p\geq
1$, is defined by $||X^n||_p = \left(\sum_i^n
|X_i|^p\right)^{\frac{1}{p}}$. A sequence $X^n$ is said to converge
in $L^p$ to $Y^n$, $0\leq p\leq \infty$, if $\lim_{n\rightarrow
\infty} {\cal E}\left[||X^n-Y^n||_p\right]= 0$. We then have the
following lemma \cite[Theorem 4.1.4]{Chung01}.

\begin{lemma}
If a sequence $X^n$ converges to another sequence $Y^n$ in $L^p$,
$0\leq p\leq \infty$, then $X^n$ converges to $Y^n$ in
probability.$\hfill \Box$ \label{lemmaLP}
\end{lemma}

We can now show the desired statement.
\begin{theorem}
The sequence $U^n$ converges to the sequence $W^n$ in probability.
\end{theorem}

{\bf Proof}: According to Lemma \ref{lemmaLP}, we only need to show
$ \lim_{n\rightarrow \infty} {\cal E}[||U^n-W^n||_4]= 0$. Here,
\begin{eqnarray}
{\cal E}[||U^n-W^n||_4] & = & {\cal E}\left[\left(\sum_{i=1}^n
|U_i-W_i|^4\right)^{\frac{1}{4}}\right] \leq \left({\cal E}\left[\sum_i^n |U_i-W_i|^4\right]\right)^{\frac{1}{4}}\nonumber\\
&=& \left(\sum_i^n {\cal
E}\left[|U_i-W_i|^4\right]\right)^{\frac{1}{4}},
\label{allupperbound}
\end{eqnarray}
For any $i = 1,\cdots , n$, we have
\begin{eqnarray}
{\cal E}\left[|U_i-W_i|^4\right] &=& \int_0^1 \Pr\left(|U_i-W_i|^4> u\right) du = \int_0^1 \Pr\left(|U_i-W_i|> u^{\frac{1}{4}}\right) du \nonumber\\
               &\leq & \int_0^1 2e^{-2nu^{\frac{1}{2}}}du,
               \label{callDKW}
\end{eqnarray}
where (\ref{callDKW}) follows from (\ref{DKW}). By letting $t={\sqrt
u}$ and integrating by parts, we show
\begin{equation}
{\cal E}\left[|U_i-W_i|^4\right] \leq 4 \int_0^1 te^{-2nt}dt
=\frac{1}{n^2}-\frac{e^{-2n}}{n}(2+\frac{1}{n})
               \leq \frac{1}{n^2}. \label{indupperbound}
\end{equation}
Combining (\ref{allupperbound}) and (\ref{indupperbound}), we obtain
\[
{\cal E}\left[||U^n-W^n||_4\right] \leq \left(\sum_{i=1}^n
\frac{1}{n^2}\right)^{\frac{1}{4}} = n^{-\frac{1}{4}},
\]
which tends to 0 as $n\rightarrow \infty$. This completes the proof
of the theorem. $\hfill \Box$

The conversion from $X^n$ (or $Y^n$) to $U^n$ (or $V^n$) can be
accomplished using a procedure that requires no computation and
relies only on a sorting algorithm. It has the important side
benefit that the output is inherently fixed-point, which is critical
in the implementation of most modern communication systems. Let $A$
be the number of bits to be used for each output sample $U_i$. This
implies that $U_i$ is of value $\frac{j}{2^A}$, $0\leq j\leq 2^A-1$.
Denote by $C(j)$, $0\leq j\leq 2^A$, the number of output samples of
value $\frac{j-1}{2^A}$. The values of $C(j)$ are determined by the
following pseudo-code:

\begin{center}
\fbox{\begin{minipage}{5.5cm}

\noindent $C(0)\leftarrow 0;$

\noindent ${\bf for}\ j=1$ to $2^A$

\noindent $\hspace{0.3in} C(j)\leftarrow \left\lfloor \frac{j\cdot
n}{2^A}\right\rfloor-\sum_{k=0}^{j-1}C(k);$

\noindent ${\bf end}$

\end{minipage}}
\end{center}

\noindent where $\lfloor x \rfloor$ is the largest integer less than
$x$. For an input sample $X_i$ with
\[
\sum_{j=0}^{k}C(j)\leq K_n(X_i)<\sum_{j=0}^{k+1}C(j),
\]
the corresponding output $U_i$ is given by $\frac{k}{2^A}$.

To efficiently implement this process, we follow a
three step process: i) sort the input samples $X^n$ in ascending order; ii)
convert sorted samples to values $\frac{j}{2^A}$, $0\leq j\leq
2^A-1$; iii) associate each input sample with its converted value.

Suppose input samples $X^n$ are sorted to $\widetilde{X}^n$, where
$\widetilde{X}_1\leq \cdots \leq \widetilde{X}_n$. The index mapping
between $X^n$ and $\widetilde{X}^n$ is also recorded for the use in
the association step.

The values of $\widetilde{X}^n$ are converted to $\widetilde{U}^n$
using the algorithm defined via the pseudo-code below. The algorithm
distributes $n$ items among $A$ bins in a ``uniform" way even when
$A$ does not divide $n$. The process is based on the rate-matching
algorithms used in modern cellular systems, e.g. \cite{3GPP}, and is
also similar to line-drawing algorithms in computer graphics.

\begin{center}
\fbox{\begin{minipage}{9cm}

\noindent $c\leftarrow 0; \hspace{0.3in} k\leftarrow 0;
\hspace{0.3in} j\leftarrow 1;$

\noindent ${\bf while}\ (j\leq n)$

\noindent\hspace{0.3in} $c\leftarrow c+\frac{n}{2^A};$

\noindent\hspace{0.3in} ${\bf while}\ (c\geq 1)$

\noindent\hspace{0.6in} ${\widetilde U}_j \leftarrow \frac{k}{2^A};
\hspace{0.3in} j\leftarrow j+1; \hspace{0.3in} c\leftarrow c-1;$

\noindent\hspace{0.3in} ${\bf end}$

\noindent\hspace{0.3in} $k\leftarrow k+1;$

\noindent ${\bf end}$
\end{minipage}}
\end{center}

The last step rearranges $\widetilde{U}^n$ to outputs $U^n$ such
that the $i^{th}$ output sample $U_i$ is associated with the
$i^{th}$ input sample $X_i$.

Although the above procedures use $2^A$ as the total number of
possible values to be assigned, in general, any integer $M$ may be
substituted for $2^A$, in which case the unit interval $[0,1)$ is
partitioned into $M$ equal sub-intervals, with the data distributed
among them as uniformly as possible.

To equiprobably quantize uniformly distributed samples $U^n$ with
$v$ bits per sample, the Quantizer determines the quantization
boundaries as
\[
q_i=\frac{i}{2^v}, \ \ \ 0\leq i\leq 2^v.
\]
For a simple decoding process, the quantization error $E$ is defined
as the difference between $U$ and the lower bound of the interval to
which $U$ belongs. Hence, the quantization error $E$ is uniformly
distributed between 0 and $\frac{1}{2^v}$. The transmission of such
quantization errors $E^n=(E_1,\cdots , E_n)$ over the public channel
does not reveal any information about ${\bf X}_b$.

For the case of fixed point inputs $U^n$, if the number of bits per
sample $v$ in the Quantizer block used for generating ${\bf X}_b$ is
less than the number $A$ of bits used for $U$, then the Quantizer
block obtains the quantized value and the quantization error for $U$
simply from the first $v$ bits and the last $A-v$ bits out of the
$A$ bits for $U$, respectively.

Bob's Data Converter performs the same operations as Alice's. The
Channel Decoder calculates the per-bit LLR based on the outputs of
Bob's Data Converter block $V^n$ and the received quantization
errors $E^n$. Unlike the jointly Gaussian model, the joint
distribution of $X$ and $Y$ in this case is unknown and the accurate
LLR is generally incomputable.

We provide an extremely simple but effective way of computing the
LLR. Heuristically, the LLR is related to the distances from $V$ to
the possible $U$ values that cause $X_{b,i}=1$ and that cause
$X_{b,i}=0$. Suppose a uniform sample $U$ is quantized and Gray
coded to bits $(X_{b,1},\cdots ,X_{b,v})$ and the quantization error
of $U$ is $E$. The heuristic LLR $L_i$ for $X_{b,i}$, $1\leq i\leq
v$, is derived through the following pseudo-code:

\begin{center}

\fbox{\begin{minipage}{6.5cm}

\noindent ${\bf for}\ i=1$ to $v$

\noindent\hspace{0.3in} $L_i\leftarrow 2E -2V +1-2^{-(v-i+1)};$

\noindent\hspace{0.3in} ${\bf if}\ V<0.5$

\noindent\hspace{0.6in} $V\leftarrow 2V;$

\noindent\hspace{0.6in} $E\leftarrow 2E;$

\noindent\hspace{0.3in} ${\bf else}$

\noindent\hspace{0.6in} $V\leftarrow 1-2V;$

\noindent\hspace{0.6in} $E\leftarrow 2^{-(v-i)} - 2E;$

\noindent\hspace{0.3in} ${\bf end}$

\noindent ${\bf end}$

\end{minipage}}

\end{center}

Consider an example of $E=0.2$ and $v=1$. This quantization error
indicates the two possible values of $U$ are 0.2 and 0.7, which
corresponds to $X_{b,1}=0$ and $X_{b,1}=1$, respectively. If
$V=0.3$, which is closer to the possible $U$ value 0.2, then it is
more likely that $X_{b,1}$ is equal to `0' and the LLR for $X_{b,1}$
should be positive. It follows from the pseudo-code above that $L_1=
0.3$. If $V=0.5$, which is closer to the possible $U$ value 0.7,
then it is more likely that $X_{b,1}$ is equal to `1' and the LLR
for $X_{b,1}$ should be negative. It follows from the codes above
that $L_1= -0.1$.

As the $L_i$ obtained in the codes above is generally within the
range of $[-1,1]$, the likelihood probability of each bit is
restricted to the range of $[0.27, 0.73]$. Hence, it is desirable to
re-scale $L_i$ to the operational range of the modified
belief-propagation algorithm by multiplying with a constant.

\subsubsection{Simulation and Experimental Validation}

We examine the performance of the proposed approach in a simulation
environment with the jointly Gaussian channel model and with real
channels.

\begin{figure}
          \centering
          \includegraphics[width=4in, height=3.2in]{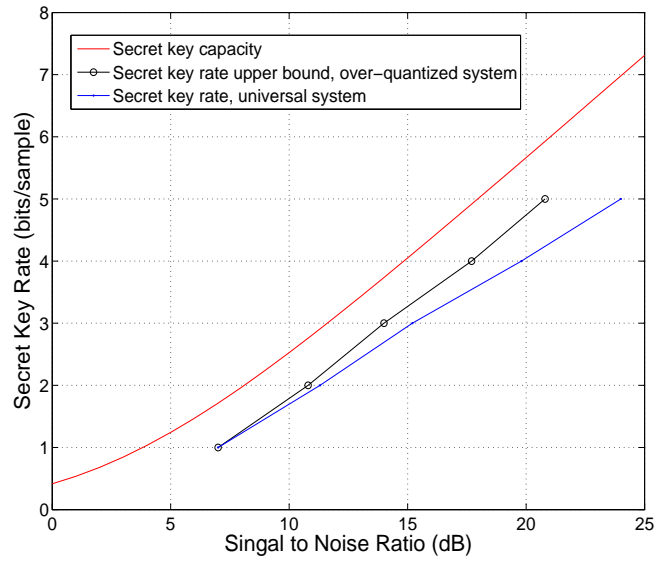}
           \caption{Secret key rates achieved by the universal system.}
           \label{fig8}
         \end{figure}

In order to examine the performance of the universal system, we
apply it to the jointly Gaussian model, though noting that the parameters $P$, $N$ of
the jointly Gaussian model are not utilized in the universal system.
The secret key rates achieved by the universal system are shown in
Figure \ref{fig8}. For comparison, the secret key capacity and the
upper bound for the secret key rates achieved by the over-quantized
system are also plotted in the same figure. It is seen from the
figure that the universal system performs well at low SNR, but
deviates at high SNR. The deviation may be due to the trade-off made
between the regularly quantized bits and the over-quantized bits. A
different trade-off can push the deviation point higher at the
expense of more communication (of over-quantized bits) and higher
LDPC decoding complexity.


We experimentally validated the feasibility of the above universal
approach using 802.11 setup described earlier. In the two
experiments stated in Section II, Bob sent Probe Request messages at
an average rate of 110 ms.\footnote{Here, we assume the channel
coherence time is less than or equal to 110 ms. Hence, two
consecutive CIRs at either terminal are assumed to be mutually
independent.} Typically, Bob received the corresponding Probe
Response message from Alice within 7 ms after a Probe Request
message was sent. It is reported in Table I that in the first
experiment, the mutual information between Alice and Bob's samples
is about 3.294 bits/sample, while the mutual information between Bob
and Eve's samples is about 0.047 bit/sample. In the second
experiment, the mutual information between Alice and Bob's samples
is about 1.218 bits/sample, while the mutual information between Bob
and Eve's samples is 0 within the accuracy of the measurement. This
suggests that the respective secret key capacities\footnote{We abuse
the notion of capacity a bit as this ``capacity'' assumes i.i.d.
channel samples.} of the first and the second experimental
environments are about 30 ($\approx$ (3.294-0.047) bits/sample
$\div$ 0.11 second/sample) bps and 11 bps, provided that the channel
coherence time is around 110 ms.

Next, we check the secret key rates achieved by the universal
system. For the purpose of generating keys in a short time duration,
we apply a LDPC code with a shorter block length in the universal
system. The code is a (3,6) regular LDPC code of codeword length 400
bits. The quantization parameter $v$ is chosen as 3 for the first
experiment and 2 for the second experiment. This implies that for
each run of the system, a block of 134 ($\approx 400/3$) first
experimental samples or 200 second experimental samples is sent to
the universal system.

Our experimental results show that in both cases, Bob is able to
successfully decode Alice's bit sequence ${\bf X}_b$ of 400 bits.
With the reduction of 200 bits, revealed as syndrome bits over the
public channel, both terminals remain with 200 secret bits. In order
to remove the correlation between the 200 secret bits and Eve's
samples in the first experiment, which shows non-zero mutual
information, we may need to squash out an additional 7 ($\approx
0.047 * 134$) bits from the 200 secret bits, resulting in 193 secret
bits. Considering the period of collecting these 134 or 200 samples,
we conclude that the secret key rate achieved by the universal
system is about 13 bps for the first experiment and 9 bps for the
second experiment.

\section{Conclusions}
The wireless medium creates the unique opportunity to exploit
location-specific and time-varying information present in the
channel response to generate information-theoretically secret
bits, which may be used as cryptographic keys in other security services. This ability follows from the property that in a
multipath scattering environment, the channel impulse response
decorrelates in space over a distance that is of the order of the
wavelength, and that it also decorrelates in time, providing a
resource for fresh randomness. In this paper, we have studied secret
key extraction, under the assumption of a Rayleigh or Rician fading
channel, and under a more general setting where we do not make any
assumption on the channel distribution. We have developed two
techniques for producing identical secret bits at either end of a
wireless communication link and have evaluated each technique using
channel measurements made using a modified 802.11 system. The first
technique is based on the observation of correlated excursions in
the measurements at the two users while the second technique employs
error-correction codes. The former method trades off the performance
of the latter with a lower complexity and does not require knowledge
of the channel coherence time. Since the time-varying nature of the
channel acts as the source of randomness, it limits the number of
random bits that can be extracted from the channel for the purpose
of a cryptographic key. The second method applies to more general distributions for the shared
channel information between a transmitter and receiver, and is able to achieve improved secret key rates at the tradeoff of increased complexity. Our evaluations indicate that typical indoor
wireless channels allow us to extract secret bits at a practically
useable rate, with minimal information about these secret bits being learned by an eavesdropper. Lastly, we note that as a final step, the legitimate participants in the protocol may wish to employ privacy amplification to provide added assurance that the eavesdropper cannot infer the bits being generated.

\begin{table}
\scriptsize{
\begin{center}
\begin{tabular}{| l|c |}
\multicolumn{2}{c}{}\\\hline Value of $m$ used & 4 \\\hline Choice
of $q_+, q_-$ & mean $\pm 0.125\sigma$   \\\hline Duration of
experiments & $1326$ sec ($\sim22$ min.) \\\hline Inter-probe
duration & $110$ msec. \\\hline \multicolumn{2}{|l|}{\textbf{Static
case:}}\\\hline
$I$(Alice; Bob)& $3.294 ~bits$   \\ \hline $I$(Bob; Eve)  & $0.047
~bits$  \\\hline \multicolumn{2}{|l|}{\textbf{Mobile case:}}\\\hline
$I$(Alice; Bob)&  $1.218 ~bits$  \\ \hline $I$(Bob; Eve)  &  $0.000
~bits$  \\\hline
\end{tabular}
\caption{\label{TableMI}\scriptsize{Mutual information (M.I.)
$I(u_1; u_2)$ between the measurements of users $u_1$ and $u_2$.}}
\end{center}}
\vspace{-0.5cm}
\end{table}




\end{document}